\pdfoutput=1

\documentclass[acmsmall,screen]{acmart}

\setcopyright{rightsretained}
\acmDOI{10.1145/3632851}
\acmYear{2024}
\copyrightyear{2024}
\acmSubmissionID{popl24main-p52-p}
\acmJournal{PACMPL}
\acmVolume{8}
\acmNumber{POPL}
\acmArticle{9}
\acmMonth{1}
\received{2023-07-11}
\received[accepted]{2023-11-07}

\usepackage{booktabs}   
\usepackage{subcaption} 
\usepackage{nameref}
\usepackage{hyperref}
\usepackage{iris}
\usepackage{includes}
\usepackage{todonotes}
\usepackage{aneris}
\usepackage{mdframed}
\usepackage[inline]{enumitem}
\usepackage{amsmath}
\usepackage{nameref}
\usepackage{wrapfig}
\usepackage{array}
\usepackage{mathrsfs}
\usepackage{prftree}
\usepackage{microtype}
\usepackage[capitalize,nameinlink,noabbrev]{cleveref}
\crefname{section}{\S\!\!}{\S\!\!}
\crefname{subsection}{\S\!\!}{\S\!\!}

\makeatletter
\newcommand\labeledtext[2]{\phantomsection\def\@currentlabel{#2}#2\label{#1}}
\makeatother

\AtEndPreamble{ 
  
  \theoremstyle{acmdefinition}
  \newtheorem{remark}[theorem]{Remark}
}

\usepackage{iris}

\allowdisplaybreaks

\begin{document}


\title{Trillium: Higher-Order Concurrent and Distributed Separation Logic for Intensional Refinement}

\author{Amin Timany}
\orcid{0000-0002-2237-851X}
\affiliation{%
  \institution{Aarhus University}
  \country{Denmark}
}
\email{timany@cs.au.dk}

\author{Simon Oddershede Gregersen}
\orcid{0000-0001-6045-5232}
\affiliation{%
  \institution{Aarhus University}
  \country{Denmark}
}
\email{gregersen@cs.au.dk}

\author{Léo Stefanesco}
\orcid{0000-0002-4719-2922}
\affiliation{%
  \institution{MPI-SWS}
  \country{Germany}
}
\email{lstefane@mpi-sws.org}

\author{Jonas Kastberg Hinrichsen}
\orcid{0000-0001-6143-9031}
\affiliation{%
  \institution{Aarhus University}
  \country{Denmark}
}
\email{hinrichsen@cs.au.dk}

\author{Léon Gondelman}
\orcid{0000-0001-8262-6397}
\affiliation{%
  \institution{Aarhus University}
  \country{Denmark}
}
\email{gondelman@cs.au.dk}

\author{Abel Nieto}
\orcid{0000-0003-2741-8119}
\affiliation{%
  \institution{Aarhus University}
  \country{Denmark}
}
\email{abeln@cs.au.dk}

\author{Lars Birkedal}
\orcid{0000-0003-1320-0098}
\affiliation{%
  \institution{Aarhus University}
  \country{Denmark}
}
\email{birkedal@cs.au.dk}

\renewcommand{\shortauthors}{Timany, Gregersen, Stefanesco, Hinrichsen,
Gondelman, Nieto, Birkedal}


\begin{abstract}
  Expressive state-of-the-art separation logics rely on step-indexing to model semantically complex features and to support modular reasoning about imperative higher-order concurrent and distributed programs.
  Step-indexing comes, however, with an inherent cost: it restricts the adequacy theorem of program logics to a fairly simple class of safety properties.

  In this paper, we explore if and how \emph{intensional refinement} is a viable methodology for strengthening higher-order concurrent (and distributed) separation logic to prove non-trivial safety and liveness properties.
  Specifically, we introduce \Trillium{}, a language-agnostic separation logic framework for showing intensional refinement relations between \emph{traces} of a program and a model.
  We instantiate \Trillium{} with a concurrent language and develop \Fairis{}, a concurrent separation logic, that we use to show liveness properties of concurrent programs under fair scheduling assumptions through a fair liveness-preserving refinement of a model.
  We also instantiate \Trillium{} with a distributed language and obtain an extension of \Aneris{}, a distributed separation logic, which we use to show refinement relations between distributed systems and  \TLA{} models.

\end{abstract}



\maketitle


\section{Introduction}
\label{sec:introduction}

There is a tension between the \emph{expressivity} of program logics and how much they say about the semantics of the programs being verified, that is, the strength of their \emph{adequacy theorems}.
As program logics become more expressive---to handle sophisticated programming language features such as higher-order functions and references, and to support modular and general library specifications---they require increasingly complex semantic models to justify them.

In this work, we consider \Iris \citep{Iris}, state of the art in terms of expressivity: in particular, it supports higher-order quantification, nested Hoare triples, higher-order ghost state, and \emph{impredicative} invariants, i.e., invariants that can contain any \Iris proposition, including invariants.
Through its adequacy theorem, the \Iris{} program logic \citep{irisjournal} is designed to prove three kinds of properties about programs: (1) postconditions, \ie{}, properties of the final values computed, (2) progress, \ie{}, the program never gets stuck, and (3) preservation of invariants, \ie{}, all invariants stated by the user hold throughout execution.
Note how these properties form a particular class of \emph{safety} properties.%
\footnote{Recall that a \emph{safety} property is a property that expresses ``nothing bad ever happens throughout execution'', as opposed to \emph{liveness} properties which express that ``something good will eventually happen''.} That is---as trace properties---these are somewhat trivial, in that they are all of the form: ``for any $s$ such that $s_0 \to^\ast s$, we have $P(s)$'' for some property $P$, where $\to^*$, is the reflexive-transitive closure of the operational semantics of the program. In particular, this class of properties does not include liveness properties or non-trivial safety trace properties like: ``the value of the counter must increase monotonically without skipping over any number.''
While proving non-trivial safety trace properties is not a fundamental limitation of the \Iris{} logic,\footnote{For instance, \citet{10.1145/3473586} do use an \Iris{} program logic to prove some limited intensional safety trace properties, \eg{}, ``a file can only be accessed if it has previously been opened and not subsequently closed.''} the lack of support for liveness properties is.
This inherent limitation is related to the fact that \Iris{}'s semantic model relies on step-indexing, which is crucial for the soundness of impredicative invariants.
This fact is a compromise that the designers of \Iris{} have made.
On the one hand, making the semantic model step-indexed, and thereby enabling impredicative invariants, allows many important applications.
Notably, it allows us to construct logical relations models for proving type safety and contextual equivalences for expressive programming languages, \eg{}, System F with recursive types, higher-order references, and concurrency \citep{ipm}.
On the other hand, step-indexing inherently restricts program logics to only be able to express properties that \emph{concern} finite prefixes of program execution.
This effectively dooms any program logic developed on top of the \Iris{} base logic from supporting liveness properties---at least directly.
The thesis that we explore in this work is that \emph{refinement is a viable methodology for strengthening higher-order concurrent (and distributed) separation logic to prove non-trivial safety and liveness trace properties.}

\paragraph{Focus and Methodology}
In this paper, we use \Iris{} to establish \emph{intensional refinements} between programs and labeled transition systems (LTSs), including the strong notion of \emph{liveness-preserving} refinements, for \emph{concurrent and distributed programs}.
We develop \Trillium{}, a language-agnostic generic program logic, whose adequacy theorem guarantees the existence of a refinement between the program and an LTS chosen by the prover.
This is in addition to the usual properties enjoyed by program logics for safety reasoning as mentioned above, \ie{}, postconditions, progress, and preservation of invariants.
The key insight is that, by showing an \emph{intensional} refinement between a program and an LTS, we can---\emph{indirectly}---establish non-trivial safety trace properties and liveness properties such as fair termination of concurrent programs.
By proving that the LTS enjoys the property of interest (which is often, if not always, simpler than proving it for the program itself), we can use the refinement relation to ``transport'' the property to the program.
For this reason, we will refer to the LTS as a \emph{model} or \emph{specification} of a program \emph{implementation}.

\pagebreak

\paragraph{Contributions}
In summary, we make the following contributions:
\begin{itemize}
\item We introduce \Trillium{} (\cref{sec:background_trillium}), a language-agnostic separation logic framework for establishing intensional refinement relations between traces of program executions and models.
\item We develop \Fairis{} (\cref{sec:fairis}), a higher-order concurrent separation logic for showing liveness properties of concurrent programs under fair scheduling assumptions through a fair termination-preserving refinement of a carefully chosen model (see discussion in \cref{sec:bg-key-obs}).
\item We showcase \Fairis{} on a number of challenging but small concurrent examples (\cref{sec:fairis_examples}).
\item We instantiate \Trillium{} with \AnerisLang{} to get an extension of \Aneris{} \cite{DBLP:conf/esop/Krogh-Jespersen20} that can be used to show intensional refinements of distributed systems (\cref{sec:aneris}).
\item We use the Aneris instantiation of Trillium to show that two distributed protocols, two-phase commit and single-decree Paxos, refine their abstract \TLA{} \cite{lamport1993hybrid} specifications.
  To the best of our knowledge, this is the first foundationally verified proof that a concrete implementation of a distributed protocol correctly implements an abstract \TLA{} specification.
\item We further show functional correctness and strong eventual consistency of a concrete implementation of a Conflict-Free Replicated Data Type (CRDT).
  The challenging part is incorporating the notion of fairness of the inter-replica communication; if messages from one replica are just ignored, then eventual consistency will never be reached.
  Moreover, the concurrent interactions with the user-exposed operations makes it non-trivial to reason about eventual consistency.
  To the best of our knowledge, this is the first such proof that takes into account the inter-replica communication at the level of the implementation.
  For the sake of space, we have relegated further details about this example to
  the accompanying appendix.
\item All the results that appear in the paper have been formalized in the Coq proof assistant using the Iris separation logic framework.
\end{itemize}

\newcommand{\forsimrel}[1]{\mathit{ForSim}_{#1}}
\newcommand{\Traces}{\mathit{tr}}
\newcommand{\TraceLast}{\mathit{last}}
\newcommand{\TraceFirst}{\mathit{first}}
\section{Background and Key Observations}
\label{sec:bg-key-obs}

We will think of the operational semantics of a concurrent program as an LTS where the transition labels are thread identifiers corresponding to the thread taking the step.
That an LTS refines another is a standard notion: two states are in a refinement relation if there exists a \emph{forward simulation relation} $R$ that relates them (see, \eg{}, \citet{CLEAVELAND2001391}).\footnote{\citet{CLEAVELAND2001391} work with relations on a single LTS whereas we work with two LTSs.
Nonetheless, all the results carry over straightforwardly by simply considering an LTS that is the disjoint union of the two LTSs we consider.}
The goal in this paper is to transport \emph{intensional} safety and liveness properties of (possibly infinite) \emph{traces} along such a refinement relation, \eg{}, transporting the property ``the value of the counter increases (or stays the same) monotonically without skipping over any number'' from the model \ref{lts:chain} in \cref{fig:chain-ltss} to the program \aneris{count_up} in \cref{fig:prog-count-up}.

To transport intensional properties, we will work with \emph{intensional refinement}, which is a lockstep relation where every step of the program is matched by a step of the model.
This is, of course, too strong if taken literally:
for example, the step of computation corresponding to a recursive call of \aneris{count_up} does not increment the counter and hence does not correspond to a step in the model.
For now, we will ignore this issue; in the following section we will present constructions on LTSs that will allow us to relax the correspondence between the program and the model, while still allowing intensional properties to be transported.

\begin{figure}
  \begin{center}
    \begin{minipage}[b]{0.45\linewidth}
      \begin{center}
        \begin{tikzpicture}[every node/.style={scale=0.7}, scale=0.7]
          \draw (0, 0) node[draw, thick, circle] (WZERO) {$\mathbf{0}$};
          \draw (1.5, 0) node[draw, thick, circle] (WONE) {$\mathbf{1}$};
          \draw (3, 0) node[draw, thick, circle] (WTWO) {$\mathbf{2}$};
          \draw (4.5, 0) node[draw, thick, circle] (WTHREE) {$\mathbf{3}$};
          \draw (6, 0) node (WDOTS) {\bfseries \Large \dots};

          \draw (WZERO) edge[-stealth, thick] (WONE);
          \draw (WONE) edge[-stealth, thick] (WTWO);
          \draw (WTWO) edge[-stealth, thick] (WTHREE);
          \draw (WTHREE) edge[-stealth, thick] (WDOTS);
        \end{tikzpicture}
      \end{center}
      \vspace{2.4em}
      \begin{center}
        \tiny\bfseries
        (\labeledtext{lts:chain}{Chain}) Chain of natural numbers
      \end{center}
    \end{minipage}
    \begin{minipage}[b]{0.5\linewidth}
      \begin{center}
        \begin{tikzpicture}[every node/.style={scale=0.7}, scale=0.7]
          \draw (8, 0) node[draw, thick, circle, inner sep=0.3ex] (ZERO_ZERO) {\small $\mathbf{(\infty, 0)}$};
          \draw (9, 1) node[draw, thick, circle, inner sep=0.3ex] (ONE_ONE) {\small $\mathbf{(1, 1)}$};
          \draw (10.75, 1) node[draw, thick, circle, inner sep=0.3ex] (TWO_ONE) {\small $\mathbf{(2, 1)}$};
          \draw (12.25, 1) node[draw, thick, circle, inner sep=0.3ex] (TWO_TWO) {\small $\mathbf{(2, 2)}$};
          \draw (10, 0) node[draw, thick, circle, inner sep=0.3ex] (THREE_ONE) {\small $\mathbf{(3, 1)}$};
          \draw (11.5, 0) node[draw, thick, circle, inner sep=0.3ex] (THREE_TWO) {\small $\mathbf{(3, 2)}$};
          \draw (13, 0) node[draw, thick, circle, inner sep=0.3ex] (THREE_THREE) {\small $\mathbf{(3, 3)}$};
          \draw (10, -1.2) node[outer sep = 0.5ex] (FOUR_DOTS) {\raisebox{1em}{\bfseries \Large \dots}};

          \draw (ZERO_ZERO) edge[-stealth, thick] (ONE_ONE);

          \draw (ZERO_ZERO) edge[-stealth, thick] (TWO_ONE);
          \draw (TWO_ONE) edge[-stealth, thick] (TWO_TWO);

          \draw (ZERO_ZERO) edge[-stealth, thick] (THREE_ONE);
          \draw (THREE_ONE) edge[-stealth, thick] (THREE_TWO);
          \draw (THREE_TWO) edge[-stealth, thick] (THREE_THREE);

          \draw (ZERO_ZERO) edge[-stealth, thick, dotted] (FOUR_DOTS);
        \end{tikzpicture}
      \end{center}
      \begin{center}
        \tiny\bfseries
        (\labeledtext{lts:finchains}{FinChains}) All finite chains of numbers; the first component is the length
      \end{center}
    \end{minipage}
  \end{center}
  \caption{Two simple LTSs representing the \emph{infinite} chain of natural numbers \ref{lts:chain}, and all \emph{finite} chains of natural numbers \ref{lts:finchains}.}
  \label{fig:chain-ltss}
\end{figure}
\begin{figure}
  \begin{center}
    \begin{minipage}{27em}
      \begin{AnerisPL}[numbers=none]
        let rec count_up l = FAA l 1; count_up l in count_up l
      \end{AnerisPL}
    \end{minipage}
  \end{center}
  \caption[The program \aneris{count_up}.]{The program \aneris{count_up}. Here \aneris{FAA} is the (atomic) fetch-and-add operation which increments the integer stored in its first argument (a reference) by the given amount in the second argument. We assume that the value of $l$ is zero at the beginning.}
  \label{fig:prog-count-up}
\end{figure}
We recall the precise definitions of forward simulation and refinement. The definitions are relative to a parameter $\xi$,
a relation on traces, which provides for a bit of flexibility, by allowing one to restrict attention to traces satisfying $\xi$.

\begin{definition}[$\xi$-forward simulation]\label{def:forward-sim}
  Let $\xi$ be a binary relation on finite traces.
  A relation $R$ is a \emph{$\xi$-forward simulation}, written $\forsimrel{\xi}(R)$, if:
  \begin{align*}
    \forsimrel{\xi}(R) \eqdef{} & \big(\forall \tau, \tau'.\; R(\tau, \tau') \implies \xi(\tau, \tau')\big) \land{}\\
    & \big(\forall \tau, \tau', l, s.\; R(\tau, \tau') \land \TraceLast(\tau) \lto{l} s \implies \exists l', s'.\; R(\tau \lto{l} s, \tau' \lto{l'} s')\big)
  \end{align*}
  where $\TraceLast$ maps a trace to its end state.
\end{definition}

\begin{definition}[Intensional refinement]\label{def:inten-ref}
  Let $\xi$ be a binary relation on finite traces.
  A finite trace $\tau$ is an \emph{intensional refinement} of a finite trace $\tau'$, with respect to parameter $\xi$, written $\tau \xisim \tau'$, if there exists a $\xi$-forward simulation relation $R$ such that $R(\tau, \tau')$.
  That is,
  \[
    \tau \xisim \tau' \eqdef{} \exists R.\; \forsimrel{\xi}(R) \land R(\tau, \tau')
  \]
\end{definition}
%
%
In our running example, the parameter $\xi$ is used to restrict our attention to traces where the value of the counter and the model state agree at all times.

The \Trillium{} program logic is designed to establish (\cf{}~\cref{thm:adequacy}) an intensional refinement $c \xisim m$ between singleton traces consisting of a program state $c$ and a model state $m$.
From \Cref{lem:extending-int-ref} below it then follows directly that any possibly-infinite execution of the program can be matched by a possibly-infinite trace of the model in such a way that all their corresponding finite prefixes are in the intensional refinement relation.

\begin{definition}[Trace Relation Extension]
  Given a relation $R$ on finite traces, we lift $R$ to possibly-infinite traces, written $\tracepredof{R}$, by considering all finite prefixes as follows:
  \[
    \tracepredof{R}(\tau_1, \tau_2) \eqdef{} \forall \tau_1', \tau_2'.\; \vert\tau_1'\vert = \vert\tau_2'\vert \land \traceprefix{}(\tau_1', \tau_1) \land \traceprefix{}(\tau_2', \tau_2) \implies R(\tau_1', \tau_2')
  \]
  where $\traceprefix(\tau, \tau')$ means $\tau$ is a prefix of $\tau'$.
\end{definition}

\begin{lemma}[Intensional refinement, lifting]
  \label{lem:extending-int-ref}
  Let $L_1$ and $L_2$ be two LTSs, and $\xi$ a relation on traces between these LTSs.
  Let $s_1 \in L_{1}$ and $s_2 \in L_{2}$ be two states such that $s_1 \xisim s_2$ (seen as singleton traces).
  For all possibly-infinite traces $\tau_{1}$ of $L_{1}$ such that
  $\TraceFirst(\tau_{1}) = s_{1}$ there exists a possibly-infinite trace
  $\tau_{2}$ where $\TraceFirst(\tau_{2}) = s_{2}$ such that
  $\tau_{1}~\tracepredof{\refines}_{\xi}~\tau_{2}$.
  Here, $\TraceFirst$ maps traces to their initial state.
\end{lemma}

\subsection{Program Steps That Do Not Correspond to Steps in the Model}
The core idea of the methodology we propose in this paper is that the model refined by the program is more abstract, and hence simpler and easier to reason about, than the program.
Thus, in general, there will be steps in the program that do not correspond to any step in the model.
One way to reconcile this with our notion of refinement is to allow for \emph{stuttering}.
That is, allow the program to take a step while the model stays in the same state.
We will support stuttering by \emph{lifting} the model into an LTS that allows stuttering.
We consider two kinds of lifting, one that is only sound for intensional safety trace properties, and one that is sound for liveness trace properties as well.
For the former, we show that it is sound in the sense that, if a program refines the \emph{lifted} model, and the \emph{original} model enjoys an intensional safety trace property, so does the program.
Similarly, the latter lifting is sound with respect to liveness trace properties.

We remark on a subtle point here, namely that \Trillium{} is a framework, which means that when we combine the base program logic of \Trillium{} with one of the two aforementioned liftings, we obtain two different program logics.
The choice of lifting presents a compromise between expressivity and simplicity of the \emph{derived} logical principles of the program logic.
In particular, for the lifting that is sound for intensional safety properties, we obtain a program logic that is conservative with respect to the ordinary program logic of \Iris{} in that all the reasoning principles of the \Iris{} program logic are still sound, and in addition one obtains simple reasoning principles that allow proving refinements (\cf{}~\cref{sec:aneris}).
For the lifting that is sound with respect to liveness properties, we obtain a program logic that is more involved but allows proving liveness properties (\cf{}~\cref{sec:fairis}).

\paragraph{Stutter-Lifting: Sound for Intensional Safety Trace Properties}
This lifting is very simple: it essentially amounts to adding \emph{self-loops}, with a special label, to all states of the LTS.
For example, the LTS \ref{lts:chain} from \cref{fig:chain-ltss} would result in the following LTS, where dotted arrows are added to support stuttering:

\begin{center}
  \begin{minipage}[b]{0.5\linewidth}
    \begin{center}
      \begin{tikzpicture}[every node/.style={scale=0.7}, scale=0.7]
        \draw (0, 0) node[draw, thick, circle] (WZERO) {$\mathbf{0}$};
        \draw (1.5, 0) node[draw, thick, circle] (WONE) {$\mathbf{1}$};
        \draw (3, 0) node[draw, thick, circle] (WTWO) {$\mathbf{2}$};
        \draw (4.5, 0) node[draw, thick, circle] (WTHREE) {$\mathbf{3}$};
        \draw (6, 0) node (WDOTS) {\bfseries \Large \dots};

        \draw (WZERO) edge[-stealth, thick] (WONE);
        \draw (WONE) edge[-stealth, thick] (WTWO);
        \draw (WTWO) edge[-stealth, thick] (WTHREE);
        \draw (WTHREE) edge[-stealth, thick] (WDOTS);

        \draw (WZERO) edge[-stealth, densely dotted, thick, loop above] (WZERO);
        \draw (WONE) edge[-stealth, densely dotted, thick, loop above] (WONE);
        \draw (WTWO) edge[-stealth, densely dotted, thick, loop above] (WTWO);
        \draw (WTHREE) edge[-stealth, densely dotted, thick, loop above] (WTHREE);
      \end{tikzpicture}
    \end{center}
    \vspace{0.4em}
    \begin{center}
      \tiny\bfseries
      (\labeledtext{lts:chain-with-loop}{Chain-Loops}) Stutter-Lifting of the LTS \ref{lts:chain} in \cref{fig:chain-ltss}
    \end{center}
  \end{minipage}
\end{center}

The program \aneris{count_up} in \cref{fig:prog-count-up} is an intensional refinement of the model \ref{lts:chain-with-loop} above if we take the parameter $\xi$ to relate the value of memory location $l$ with the state of the model---the recursive call then corresponds to taking the self-loop in the relevant state.
As mentioned above, this stutter-lifting is sound for intensional safety trace properties.
Hence, to show that ``the value of the counter increases (or stays the same) monotonically without skipping over any number'', it suffices to show that this property is enjoyed by \ref{lts:chain} in \cref{fig:chain-ltss}.

However, this lifting is not sound for liveness properties, because it is also refined by (using the same $\xi$ parameter) the following program:
\begin{center}
  \begin{minipage}{27em}
    \begin{AnerisPL}[numbers=none]
      let rec loop () = loop () in FAA l 1; loop ()
    \end{AnerisPL}
  \end{minipage}
\end{center}
which only increments $l$ once and afterwards loops forever.
To see this, take a very simple liveness property like ``the value of the counter is eventually 3'' which is trivially true for the LTS \ref{lts:chain} in \cref{fig:chain-ltss}, but not for the program above.
The culprit here is unrestricted stuttering.

\paragraph{Fin-Stutter-Lifting: also Sound for Liveness Properties}
To obtain soundness with respect to liveness properties, we define a fin-stutter-lifting construction, which only allows for finite stuttering.
That is, instead of adding loops, it essentially adds finite unrollings of loops, by creating copies of each state, each of which allows at most a certain, fixed number of stuttering steps.
For instance, the Fin-Stutter-Lifting of the LTS \ref{lts:chain} in \cref{fig:chain-ltss} is given blow:
\begin{center}
  \begin{minipage}[b]{0.5\linewidth}
    \begin{center}
      \begin{tikzpicture}[every node/.style={scale=0.55}, scale=0.65]
        \draw (0, 0) node[draw, thick, circle] (WZERO) {$\mathbf{0}$};
        \draw (1, 1.7) node[draw, thick, circle, inner sep = 0.2em] (WZERO_ONE) {$\mathbf{0}_0$};
        \draw (1, 0.7) node[draw, thick, circle, inner sep = 0.2em] (WZERO_TWO) {$\mathbf{0}_1$};
        \draw (1, -0.7) node[draw, thick, circle, inner sep = 0.2em] (WZERO_THREE) {$\mathbf{0}_2$};
        \draw (1, -1.7) node[inner sep = 0.1em] (WZERO_DOTS) {$\vdots$};
        \draw (2, 0) node[draw, thick, circle] (WONE) {$\mathbf{1}$};
        \draw (3, 1.7) node[draw, thick, circle, inner sep = 0.2em] (WONE_ONE) {$\mathbf{1}_0$};
        \draw (3, 0.7) node[draw, thick, circle, inner sep = 0.2em] (WONE_TWO) {$\mathbf{1}_1$};
        \draw (3, -0.7) node[draw, thick, circle, inner sep = 0.2em] (WONE_THREE) {$\mathbf{1}_2$};
        \draw (3, -1.7) node[inner sep = 0.1em] (WONE_DOTS) {$\vdots$};
        \draw (4, 0) node[draw, thick, circle] (WTWO) {$\mathbf{2}$};
        \draw (5, 1.7) node[draw, thick, circle, inner sep = 0.2em] (WTWO_ONE) {$\mathbf{2}_0$};
        \draw (5, 0.7) node[draw, thick, circle, inner sep = 0.2em] (WTWO_TWO) {$\mathbf{2}_1$};
        \draw (5, -0.7) node[draw, thick, circle, inner sep = 0.2em] (WTWO_THREE) {$\mathbf{2}_2$};
        \draw (5, -1.7) node[inner sep = 0.1em] (WTWO_DOTS) {$\vdots$};
        \draw (6, 0) node[draw, thick, circle] (WTHREE) {$\mathbf{3}$};
        \draw (7, 1.7) node[draw, thick, circle, inner sep = 0.2em] (WTHREE_ONE) {$\mathbf{3}_0$};
        \draw (7, 0.7) node[draw, thick, circle, inner sep = 0.2em] (WTHREE_TWO) {$\mathbf{3}_1$};
        \draw (7, -0.7) node[draw, thick, circle, inner sep = 0.2em] (WTHREE_THREE) {$\mathbf{3}_2$};
        \draw (7, -1.7) node[inner sep = 0.1em] (WTHREE_DOTS) {$\vdots$};
        \draw (8, 0) node (WDOTS) {\bfseries \Large \dots};

        \draw (WZERO) edge[-stealth, thick] (WONE);
        \draw (WZERO) edge[-stealth, densely dotted, thick] (WZERO_ONE);
        \draw (WZERO_ONE) edge[-stealth, thick] (WONE);
        \draw (WZERO) edge[-stealth, densely dotted, thick] (WZERO_TWO);
        \draw (WZERO_TWO) edge[-stealth, thick] (WONE);
        \draw (WZERO_TWO) edge[-stealth, densely dotted, thick] (WZERO_ONE);
        \draw (WZERO) edge[-stealth, densely dotted, thick] (WZERO_THREE);
        \draw (WZERO_THREE) edge[-stealth, thick] (WONE);
        \draw (WZERO_THREE) edge[-stealth, densely dotted, thick] (WZERO_TWO);
        \draw (WZERO) edge[-stealth, densely dotted, thick] (WZERO_DOTS);
        \draw (WZERO_DOTS) edge[-stealth, thick] (WONE);
        \draw (WZERO_DOTS) edge[-stealth, densely dotted, thick] (WZERO_THREE);

        \draw (WONE) edge[-stealth, thick] (WTWO);
        \draw (WONE) edge[-stealth, densely dotted, thick] (WONE_ONE);
        \draw (WONE_ONE) edge[-stealth, thick] (WTWO);
        \draw (WONE) edge[-stealth, densely dotted, thick] (WONE_TWO);
        \draw (WONE_TWO) edge[-stealth, thick] (WTWO);
        \draw (WONE_TWO) edge[-stealth, densely dotted, thick] (WONE_ONE);
        \draw (WONE) edge[-stealth, densely dotted, thick] (WONE_THREE);
        \draw (WONE_THREE) edge[-stealth, thick] (WTWO);
        \draw (WONE_THREE) edge[-stealth, densely dotted, thick] (WONE_TWO);
        \draw (WONE) edge[-stealth, densely dotted, thick] (WONE_DOTS);
        \draw (WONE_DOTS) edge[-stealth, thick] (WTWO);
        \draw (WONE_DOTS) edge[-stealth, densely dotted, thick] (WONE_THREE);

        \draw (WTWO) edge[-stealth, thick] (WTHREE);
        \draw (WTWO) edge[-stealth, densely dotted, thick] (WTWO_ONE);
        \draw (WTWO_ONE) edge[-stealth, thick] (WTHREE);
        \draw (WTWO) edge[-stealth, densely dotted, thick] (WTWO_TWO);
        \draw (WTWO_TWO) edge[-stealth, thick] (WTHREE);
        \draw (WTWO_TWO) edge[-stealth, densely dotted, thick] (WTWO_ONE);
        \draw (WTWO) edge[-stealth, densely dotted, thick] (WTWO_THREE);
        \draw (WTWO_THREE) edge[-stealth, thick] (WTHREE);
        \draw (WTWO_THREE) edge[-stealth, densely dotted, thick] (WTWO_TWO);
        \draw (WTWO) edge[-stealth, densely dotted, thick] (WTWO_DOTS);
        \draw (WTWO_DOTS) edge[-stealth, thick] (WTHREE);
        \draw (WTWO_DOTS) edge[-stealth, densely dotted, thick] (WTWO_THREE);

        \draw (WTHREE) edge[-stealth, thick] (WDOTS);
        \draw (WTHREE) edge[-stealth, densely dotted, thick] (WTHREE_ONE);
        \draw (WTHREE_ONE) edge[-stealth, thick] (WDOTS);
        \draw (WTHREE) edge[-stealth, densely dotted, thick] (WTHREE_TWO);
        \draw (WTHREE_TWO) edge[-stealth, thick] (WDOTS);
        \draw (WTHREE_TWO) edge[-stealth, densely dotted, thick] (WTHREE_ONE);
        \draw (WTHREE) edge[-stealth, densely dotted, thick] (WTHREE_THREE);
        \draw (WTHREE_THREE) edge[-stealth, thick] (WDOTS);
        \draw (WTHREE_THREE) edge[-stealth, densely dotted, thick] (WTHREE_TWO);
        \draw (WTHREE) edge[-stealth, densely dotted, thick] (WTHREE_DOTS);
        \draw (WTHREE_DOTS) edge[-stealth, thick] (WDOTS);
        \draw (WTHREE_DOTS) edge[-stealth, densely dotted, thick] (WTHREE_THREE);
      \end{tikzpicture}
    \end{center}
    \begin{center}
      \tiny\bfseries
      (\labeledtext{lts:chain-with-fin-stutter}{Chain-Fin-Stutter}) Fin-Stutter-Lifting of the LTS \ref{lts:chain} in \cref{fig:chain-ltss}
    \end{center}
  \end{minipage}
\end{center}
The idea is to add states $\set{n_i \mid i \in \nat}$ (for each state $n$), all of which intuitively correspond to state $n$.
From a state $n$, we can either go to the state $n+1$, or \emph{stutter} to a state $n_s$ from where we can at most stutter $s$ times before going to state $n+1$.
The Fin-Stutter-Lifting construction is sound for liveness properties.
In fact, it is the core idea of the so-called fuel construction we present in \cref{sec:fairis_adequacy}.
Note that, as expected, the program above which increments the counter only once and then loops forever does not refine the LTS \ref{lts:chain-with-fin-stutter}.

\subsection{Step-Indexing and Finite Approximability}
As mentioned in \cref{sec:introduction}, step-indexing prevents direct reasoning about liveness properties because it restricts reasoning to finite prefixes of program execution.
The same complication arises when establishing refinement relations.
In this work, this issue shows up in \cref{thm:adequacy} as the \emph{relative image-finiteness} side-condition that we will discuss and formally define here.
In order to see this issue concretely, revisit the LTS \ref{lts:finchains} from \cref{fig:chain-ltss}.
\ref{lts:finchains} has no infinite paths and hence any program that refines the fin-stutter-lifting of \ref{lts:finchains}, regardless of the $\xi$ parameter, must terminate.
However, in a step-indexed logic, one can show a refinement relation between the program \aneris{count_up} in \cref{fig:prog-count-up} and the fin-stutter-lifting of \ref{lts:finchains}.
To see this, simply take the $\xi$ parameter for this step-indexed refinement relation to require that the value of \aneris{l} corresponds to the second component of the state in the fin-stutter-lifting of \ref{lts:finchains}.
In this case, for any finite trace of the program, there is a trace in the fin-stutter-lifting of \ref{lts:finchains} that matches it according to this $\xi$ relation.
The key point here is that when the value of \aneris{l} goes from 0 to 1 in the program, on the model side we go from the state $(\infty, 0)$ to $(n, 1)$, where $n$ is the number of steps of execution being considered.
Hence, it would not be sound, \emph{in this case}, to conclude an intensional refinement relation from the refinement relation established in the step-indexed logic.
The crux of the issue here is the unbounded choice of transitions going out of the state $(\infty, 0)$ which allows us to pick a path based on the number of steps of execution that we are considering.
Below we analyze this problem more formally, and conclude that a so-called \emph{relative image-finiteness} side-condition suffices to circumvent the problem.
(Note that the problem of unbounded branching is already present in the fin-stutter-lifting construction; we will discuss this issue further when we explain the fuel construction in \cref{sec:fairis_adequacy}.)

We first define a notion of \emph{finite approximation}, which intuitively corresponds to what \emph{guarded recursive predicates} compute.
Guarded recursive predicates are those defined as fixed points using the step-indexing technique, \eg{}, the weakest preconditions underlying the program logic of both \Iris{} and \Trillium{}.
\begin{definition}
  Let $F$ be a function on the space of binary relations on traces of LTSs $L_1$ and $L_2$, \ie{}, $F : 2^{\Traces(L_1) \times \Traces(L_2)} \to 2^{\Traces(L_1) \times \Traces(L_2)}$, where $2^A$ is the powerset of $A$, and $\Traces(L)$ is the set of all traces of the LTS $L$.
  We define the \emph{finite approximation} of $F$, written $\finapprox(F)$, as
  \begin{align*}
    \finapprox(F) \eqdef{} \bigcap_{i = 0}^{\infty} R_F^i
    \hspace{2em} && \text{ where } R_F^0 \eqdef{} \Traces(L_1) \times \Traces(L_2) \text{ and } R_F^{i + 1} \eqdef{} F(R_F^i)
  \end{align*}
\end{definition}
The upshot of the limitation of step-indexed logics is that the best we can hope to conclude from a refinement relation defined in a step-indexed logic is a finite approximation of the refinement relation.
Specifically, in our case we can conclude $\finapprox(\intreffunc)$, where $\intreffunc$ is the function whose greatest fixed point (by the Knaster-Tarski fixed point theorem) is the intensional refinement relation in \cref{def:inten-ref}:
\begin{align*}
  \intreffunc(R) = \set{(\tau, \tau') \middle | \forall l, s.\; \TraceLast(\tau) \lto{l} s \implies \exists l', s'.\; R(\tau \lto{l} s, \tau' \lto{l'} s') } \cap \xi
\end{align*}
In other words, we can only conclude $\xisim$ if $\xisim$ is finitely approximable, \ie{}, if $\xisim = \finapprox(\intreffunc)$, which is well-known not to be the case in general for refinement relations.
Indeed, one frequently used condition for finite approximation is so-called image-finiteness \citep[Thm. 2.6]{CLEAVELAND2001391}.
An LTS is said to be image-finite if, for any state $s$ and label $l$, there exists only finitely many states $s'$ such that $s \lto{l} s'$.
As we will discuss in \cref{sec:example_choosenat}, however, it is desirable to consider LTSs with infinite branching.
For this reason we relax the image-finiteness condition by considering the weaker \emph{relative image-finiteness} condition (a property of the $\xi$ relation, not the LTS), defined below.
For relative image-finiteness, it is sufficient that for any state $s$ of the program and for any transition $s \lto{l}s'$, there are only finitely many LTS transitions that correspond to $s \lto{l} s'$ allowed by the $\xi$ relation.
\begin{definition}[Relative image-finiteness]
  \label{def:rel-img-fin}
  Let $\xi$ be a relation on traces of two LTSs.
  The relation $\xi$ is \emph{relatively image-finite} if, for any $\tau$, $\tau'$, and transition $\TraceLast(\tau) \lto{l} s$, the following set is finite:
  \[\set{(l', s')~\middle|~\xi(\tau \lto{l} s, \tau' \lto{l'} s')}\]
\end{definition}
\begin{theorem}[Finite Approximation]
  Let $\xi$ be a relatively image-finite relation on traces of two LTSs.
  The intensional refinement relation $\xisim$ is \emph{finitely approximable}, \ie{}, $\xisim = \finapprox(\intreffunc)$.
\end{theorem}

\subsection{Further Discussions}
\paragraph{Coming up with the Appropriate Model}
One natural question regarding the methodology that we present in this paper is ``how does one come up with the appropriate model and the parameter $\xi$ for the verification task at hand?''
We argue that coming up with the appropriate model and $\xi$ parameter is of the same nature, and indeed part of picking the appropriate specification, \eg{}, relevant preconditions and postconditions.
Hence, there is no obvious, one-size-fits-all answer.
Indeed, the model and the relation often need to be designed so as to facilitate establishing the safety trace property or liveness property of the program we wish to prove, \eg{}, the relation $\xi$ and the model \ref{lts:chain} in \cref{fig:chain-ltss} that we chose for the program \aneris{count_up} in \cref{fig:prog-count-up}, in order to establish that ``the value of the counter increases monotonically without skipping over any number''.
Moreover, at a technical level, the limitation of relative-image-finiteness of the $\xi$ relation restricts us in the choice of the model and the $\xi$ parameter.
For these reasons, the chosen model may not be arbitrarily abstract.
It must reflect some of the core characteristics of the program, at least to the extent necessary for the trace property in question, and for the $\xi$ to be relative image-finite.
It is our hypothesis that coming up with appropriate models and $\xi$ relations is feasible in most, if not all, interesting examples.
However, in the present paper we only present the foundation and methodology of using intensional refinement to strengthen the expressivity of step-indexed higher-order concurrent (and distributed) separation logics, and support it with simple examples.
A proper experimental evaluation of the hypothesis, using a wide range of more advanced examples, is beyond the scope of the current work, and we leave it for future work.

\paragraph{What about Transfinite \Iris{}?}
Transfinite Iris \citep{transfinite-iris} is a variant of Iris whose model is step-indexed over an arbitrary ordinal (as opposed to the natural numbers used in the model of ordinary \Iris{}).
The upshot of this change is that Transfinite \Iris{} satisfies the so-called ``existential property'' \citep{transfinite-iris} which in effect renders the (relative) image-finiteness side-condition unnecessary.
We believe that the work of this paper could also be carried out on top of Transfinite \Iris{}, dispensing with the relative image-finiteness side-condition, albeit at the cost of other possible complications in proofs.%
\footnote{The compromise that Transfinite \Iris{} makes is that it no longer validates some of the basic reasoning principles regarding step-indexing, \ie{}, rules regarding interaction of the later modality with other connectives of the logic.
  It is partly due to this limitation that we opt to base our work on ordinary \Iris{} at the cost of the relative image-finiteness side-condition.}
\citet{transfinite-iris} prove termination and termination-preserving refinements of sequential programs but do not show any (preservation of) liveness properties beyond termination nor do they treat concurrent programs.

\paragraph{What about the Approaches for Contextual Refinement?}
There have been several works on establishing \emph{contextual} refinement in \Iris{} for complex sequential and concurrent programming languages \citep{ipm, reloc-conf, Timany2018ST, 10.1145/3009837.3009877, Timany:2019:MRV:3352468.3341709, 10.1145/3434288, Frumin2020ReLoCRA, 10.1145/3527318, 10.1145/3527326, DBLP:journals/corr/abs-2301-10061}.
In essence, contextual refinement boils down to showing that if one program terminates, so should the other.
In these previous works, it is established by the use of invariants in \Iris{}.
This implies that the approach only allows one to show that for any finite prefix of execution of the first program (the implementation side) there exists a finite execution of the second program (the specification side) and the final states correspond.
This notion of refinement is too weak for our purposes: (1) it says nothing about infinite executions and hence does not help us establish liveness properties, and (2) it does not allow us to transfer non-trivial safety trace properties.
In the Appendix we discuss an illustrative example.

\section{\Trillium: A trace program logic framework}
\label{sec:background_trillium}
In this section, we give a more detailed account of the general \Trillium~logic and a formal statement of its adequacy theorem.
We first detail how we instrument the operational semantics of the domain programming language with ``locale'' transition identifiers (essentially thread id's), to facilitate thread-level properties such as fair scheduling.
Second, we present the fundamentals of the \Trillium{} logic, and its adequacy theorem~(\cref{thm:adequacy}).
We focus on the novelties of \Trillium{} but will recount necessary constructions of the Iris base logic briefly at a high level.

\paragraph{Language Agnostic Framework.}
\label{sec:opsem}

The \Trillium~program logic is language agnostic and is defined with respect to any programming language which comes with an operational semantics given by a notion of expression $\expr \in \Expr$, value $\val \in \Val \subseteq \Expr$, evaluation context $\elctx \in \Ectx$, program state $\progstate \in \State$ (a model of, \eg{}, the heap and/or the network), and a \emph{primitive reduction relation} $\expr_{1}, \progstate_{1} \hstep{} \expr_{2}, \progstate_{2}; {\expr_f}_1, \cdots, {\expr_f}_n$ that relates an expression $\expr_{1}$ and a state $\progstate_{1}$ to an expression $\expr_{2}$, a state $\progstate_{2}$, and a (possibly empty) list ${\expr_f}_1, \ldots, {\expr_f}_n$ of expressions, corresponding to the threads forked by the reduction.
A value denotes an expression that has reached its final form and will no longer reduce.
We write~$\elctx[\expr]$ for the result of replacing the hole in evaluation context~$\elctx$ with~$\expr$.

The global state of the system is a \emph{configuration}~$\conf = (\threadpool, \progstate)$, where the \emph{thread pool}~$\threadpool$ is a finite mapping from \emph{locales}\footnote{The name is inspired by the Chapel programming language, where it denotes the abstract place where programs execute.} to expressions, each corresponding to an execution thread.
We will write $\threadpool(\zeta)$ for the expression whose locale is~$\zeta$ in $\threadpool$ and use $\threadpool[\zeta \mapsto \expr]$ for the corresponding update (which adds a new thread if $\zeta \not\in \dom(\threadpool)$).
We will write $\set{\zeta \mapsto \expr}$ for the singleton thread-pool consisting of a single thread $\expr$.
For a language with shared-memory concurrency a locale would simply be a thread identifier.
For a distributed language, a natural definition would be a pair~$(n, \tid)$ of the name~$n$ of the node and the thread identifier~$\tid$ of the thread in that node.
Having explicit locales as part of the language definition will be beneficial when expressing, \eg{}, thread-level properties such as fair scheduling.

The primitive reduction relation is lifted to an LTS by a relation between configurations labeled by the step-taking locale as follows:
\newcommand{\freshlocale}[2]{\mathit{fr}(#1, #2)}
\[
  \inferrule
  {(e_1, \progstate_1) \hstep{} (e_2, \progstate_2; \expr_{f_1}, \cdots, \expr_{f_n}) \and \freshlocale{\zeta}{\threadpool} = \zeta_1, \zeta_2, \ldots }
  {(\threadpool[\zeta \mapsto \elctx[\expr_{1}]], \progstate_1) \lto{\zeta} (\threadpool[\zeta \mapsto K[\expr_2], \zeta_1 \mapsto \expr_{f_1}, \ldots, \zeta_n \mapsto \expr_{f_n}], \progstate_2)}
  {}
\]
where $\freshlocale{\zeta}{\threadpool}$ is an infinite sequence of fresh locales (not in $\dom(\threadpool)$) derived from $\zeta$, \eg{}, in case of distributed systems it would consist of fresh thread identifiers on the same node as $\zeta$.

\paragraph{The \Trillium Program Logic}
\label{sec:trillium}

The goal of the \Trillium~program logic is to establish an intensional refinement $\conf \continuedsim{\progmodrel} \mstate$ between singleton traces consisting of the initial program configuration $\conf$ and a model state $\mstate$.
Proving this refinement is a matter of proving that $\xi$ \emph{always holds}, throughout the execution of the program starting in $\conf$, alongside a corresponding traversal of the model starting from $\mstate$.
This can be achieved by ensuring that the program execution makes \emph{progress} in tandem with the model, in addition to the relation $\xi$ being \emph{preserved} throughout any such execution.

Conventional Iris-style weakest precondition predicates $\wpre{\expr}[\mask]{\pred}$ guarantee \emph{postcondition validity} of terminated programs, \emph{progress} of program executions, and \emph{preservation} of the invariants whose names are in the \emph{mask} $\mask$.
This can be seen explicitly in its definition; a guarded fixpoint of the following equation \citep{irisjournal} (which, unlike in \Trillium{}, does not consider locales):
\begin{align*}
  \wpre{\expr}[\mask]{\Phi} \eqdef{}
  \big( & \expr \in \Val \asts \pred(\expr) \big) \lor{} && \textit{(post condition)} \\
  \big( & \expr \not\in \Val \asts \All \progstate . S(\progstate) \wand \pvs[\mask][\emptyset] \reducible(\expr, \progstate) \asts && \textit{(progress)} \\
        & \later \All \expr', \progstate', {\expr_{f}}_1, \cdots, {\expr_{f}}_n. (\expr, \progstate) \hstep{} (\expr', \progstate'; \expr_{f_1}, \ldots, \expr_{f_n}) \wand \\
        & \quad \pvs[\emptyset][\mask] S(\progstate') \asts \wpre{\expr'}[\mask]{\Phi} \asts \Sep_{1 \le i \le n} \wpre{\expr_{f_i}}[\top]{\Psi} && \textit{(preservation)}
\end{align*}
The definition is by case distinction: either $\expr$ is a value, in which case the postcondition should hold, or $\expr$ is not a value, in which case there are two requirements (ignoring $\pvs[\mask_1][\mask_2]$ and $S(\progstate)$ for now).
First, for the current state (captured by $S(\progstate)$), the program should be \emph{reducible}, \ie{}, it \emph{can} make progress.
Second, for any program $\expr'$ and forked threads $\expr_{f_{i}}$ that $\expr$ might reduce to, the weakest precondition must hold as well (with some post condition $\Psi$ for the forked threads).
The \emph{later modality} $\later$ guarantees that the fixpoint is well-defined (the recursive occurrence is guarded).
The \emph{update modality} $\pvs[\mask_{1}][\mask_{2}] \prop$ enables a form of rely-guarantee reasoning regarding invariants: to establish $\prop$ the prover can access the invariants in $\mask_{1}$ but they must also establish all the invariants in $\mask_{2}$ alongside proving $\prop$.
Hence, the definition of the weakest precondition preserves invariants in $\mask$ by giving access to all of them (by going from $\mask$ to $\emptyset$) but requires them to be preserved by asking them to be closed immediately after each program step (by going back from $\emptyset$ to $\mask$).
The predicate $S : \State \to \iProp$ is the \emph{state interpretation} predicate that reflects the state (\eg{}, the heap) of the program as resources in the logic and gives meaning to, \eg{}, the traditional separation logic connective $\loc \mapsto \val$ for heap ownership.
Note how the definition of weakest precondition enforces that the state interpretation, just like invariants, is preserved throughout program execution.
We often write
$\twpre{\expr}[\mask]{\Ret \val. \propB} \triangleq
\twpre{\expr}[\mask]{\Lam \val. \propB}$, and
$\twpre{\expr}[\mask]{\propB} \triangleq
\twpre{\expr}[\mask]{\Ret \val. \val = \TT \ast \propB}$.

\vspace{1em}
\begin{mdframed}
\begin{remark}[Invariants and ghost resources in \Iris{}]
  That $\prop$ is invariant in Iris is represented by the proposition $\smash{\knowInv{\namesp}{\prop}}$ which is annotated with a \emph{name} $\namesp$ that identifies it.
  In order to work with invariants formally in \Iris{}, the update modality is annotated with two masks: $\pvs[\mask_1][\mask_2]$.
  We write $\pvs[\mask]$ when $\mask_{1} = \mask_{2} = \mask$ and $\pvs$ when $\mask = \top$, the set of all masks.
  The update modality allows us to update ghost resources as described by \Iris{}'s ghost resource theory \citep{irisjournal} and to access invariants.
  Intuitively, the proposition $\pvs[\mask_{1}][\mask_{2}] \prop$ holds if we can establish $\prop$ and all invariants in $\mask_2$ through ghost updates, without violating the environment's resources (a ``frame preserving update''), and all the invariants in $\mask_1$.
  For weakest preconditions (in both the \Iris{} and \Trillium{} program logic) we can manipulate resources and invariants throughout the proof because weakest preconditions are closed under the update modality: $\pvs[\mask] \wpre{\expr}{\pred} \provesIff \wpre{\expr}{\pred} \provesIff \wpre{\expr}{\var.\; \pvs[\mask] \pred(\var)}$.
  The following rules allow us to create and access invariants and to manipulate the update modality.
  Note how accessing an invariant only makes its contents available one step later (under $\later$); similarly we only need to prove it one step later to (re)establish the invariant.
  The \ruleref{inv-access} rule---alongside providing the contents of the invariant---also tells us how we can close/reestablish the invariant.
  %
  \begin{mathpar}
  \inferH{Inv-alloc}{}{\later\prop \proves \pvs_{\mask} \knowInv{}{\prop}}
  \and
  \inferH{inv-access}
  {\namesp \in \mask}
  {\knowInv\namesp \prop \proves
    \pvs[\mask][\mask\setminus\namesp]
    \later \prop \ast
    (\later\prop \wand \pvs[\mask\setminus\namesp][\mask] \TRUE)}
  \and
  \inferH{upd-mono}
  {\prop \proves \propB}
  {\pvs[\mask_1][\mask_2] \prop \proves \pvs[\mask_1][\mask_2] \propB}
  \and
  \inferH{upd-mask-trans}{\pvs[\mask_1][\mask_2] \pvs[\mask_2][\mask_3] \propB}{\pvs[\mask_1][\mask_3] \propB}
  \and
  \inferH{upd-mask-weaken}{\mask_2 \subseteq \mask_1 \and \pvs[\mask_1][\mask_3] \propB}{\pvs[\mask_1][\mask_2] \pvs[\mask_2][\mask_3] \propB}
\end{mathpar}
These rules, along with the definition of weakest preconditions above allow us to prove a rule for accessing invariants during an ``atomic'' step of computation, \ie{}, a program that reduces to a value in a single step of computation.
\begin{mathpar}
\inferH{wp-atomic}{\namesp \in \mask \and \knowInv{\namesp}{\prop} \and \later \prop \wand \wpre{\expr}[\mask \setminus \set{\namesp}]{\var.\; \later \prop \ast \pred(v)} \and \text{atomic}(\expr)}{ \wpre{\expr}[\mask]{\pred}}
\end{mathpar}
Note how \ruleref{wp-atomic} might appear to \emph{consume} the invariant.
This is not an issue, though, as invariants are \emph{persistent} and hence freely duplicable and shareable \citep{irisjournal}.

In the rest of this paper we will make heavy use of one kind of ghost resource, which we explain in terms of abstract predicates.
Given a set $A$ we define two predicates $\authfull_\gname(a)$ and $\authfrag_\gname(a)$, for any $a \in A$, respectively called the full part and the fragment of the resource; $\gname$ is the name of the resource instance used for disambiguation.
These predicates are defined in terms of \Iris{} ghost resources internally and satisfy the following rules, which essentially say that the full part and the fragment must always agree:
\begin{mathpar}
  \inferH{auth-agree}{}{\authfull_\gname(a) \ast \authfrag_\gname(b) \proves a = b}
  \and
  \inferH{auth-alloc}{}{\proves \pvs[\mask] \exists \gname.\; \authfull_\gname(a) \ast \authfrag_\gname(a)}
  \and
  \inferH{auth-update}{}{\authfull_\gname(a) \ast \authfrag_\gname(a) \proves \pvs[\mask] \authfull_\gname(b) \ast \authfrag_\gname(b)}
\end{mathpar}
We refer to \citet{irisjournal} for a more thorough treatment of how invariants, the update modality, the later modality, and ghost state is constructed in \Iris{}.

\end{remark}
\end{mdframed}
\vspace{1em}

To define the \Trillium{} program logic, we enrich the \Iris-style weakest precondition to consider (1) program traces, (2) locales, and (3) a lock-step relation between the program trace and the model trace.
Formally, this is defined as a guarded fixpoint of the equation below:
\begin{align*}
  \twpre{&\expr}[\mask]{\pred}[\zeta] \eqdef{}\\
  &\big( \expr \in \Val \asts {\wphighlight \forall \finprogtrace, \finmodtrace.\; \stateinterp(\finprogtrace, \finmodtrace) \wand \pvs[\mask] \stateinterp(\finprogtrace, \finmodtrace)  \ast \pred(\expr)} \big) \lor{} \\
  &\big( e \not\in \Val \asts {\wphighlight \All \finprogtrace, \finmodtrace, \progstate, \elctx, \threadpool.} {\wphighlight \tracelast(\finprogtrace) = (\threadpool[\zeta \mapsto \elctx[\expr]], \progstate)
    \asts \stateinterp(\finprogtrace, \finmodtrace)} \wand \pvs[\mask][\emptyset] \\
  &\qquad \reducible(\expr, \progstate) \asts \later \All \expr', \progstate', {\expr_{f}}_1, \cdots, {\expr_{f}}_n . {(\expr, \progstate) \hstep{} (\expr', \progstate'; {\expr_{f}}_1, \cdots, {\expr_{f}}_n)} \wand \pvs[\emptyset][\mask] \\
  &\qquad\quad {\wphighlight \Exists \mstate \in \genmod, \rho.\; \tracelast(\finmodtrace) \lto{\rho} \mstate \asts}\\
  & \qquad\quad {\wphighlight
   \stateinterp(\finprogtrace \lto{\zeta} (\threadpool[\zeta \mapsto \elctx[\expr'], \freshlocale{\zeta}{\threadpool}_1 \mapsto {\expr_{f}}_1, \ldots, \freshlocale{\zeta}{\threadpool}_n \mapsto {\expr_{f}}_n], \progstate'), \finmodtrace \lto{\rho} \mstate)} \asts \\
  &\qquad\quad \twpre{\expr'}[\mask]{\pred}[\zeta] \asts {\Sep_{1 \le i \le n}} \twpre{{\expr_{f}}_i}[\mask]{\Psi}[\freshlocale{\zeta}{\threadpool}_i] \big)
\end{align*}
Instead of just program configurations, the definition now considers all program traces $\finprogtrace$ where the expression $\expr$ is about to make a step at the locale $\zeta$, under an evaluation context $\elctx$; and for all steps that $\expr$ may take, there must exist a model state $\mstate \in \genmod$ that the last state of the model trace $\finmodtrace$ can step to.
Moreover, instead of a state interpretation, it tracks a \emph{trace interpretation} $S$ of the program and model traces, which will allow us not only to interpret the state of the language and the current model state as resources.
Note also that in the case where $\expr$ is a value, in order to establish the postcondition, we can also access the trace interpretation.
This change is required for the soundness of some of our proof rules which apply even when the program is a value but whose correctness relies on the resources in the trace interpretation---we will discuss this later on.

That the definition of the Trillium weakest precondition has the intended meaning is the content of the following general adequacy theorem.



\begin{theorem}[Adequacy]\label{thm:adequacy}
  Let $\expr$ be a program, $\progstate$ a program state, $\zeta$ the locale of $\expr$ in an otherwise empty thread pool, and $\pred$ an \Iris{} predicate on values.
  Let $\mstate \in \model$ be a model state and $\progmodrel$ a relative image-finite relation on finite traces of the program and the model.
  Let $c = (\set{\zeta \mapsto \expr}, \progstate)$ be the initial configuration of the program.
  If $\xi(c, \singletontrace{\mstate})$ holds for the initial singleton traces, and furthermore we have
  \begin{align*}
    \pvs[\top]
      \stateinterp(c, \singletontrace{\mstate}) \ast \twpre{\expr}[\top]{\pred}[\zeta] \ast \alwaysholds(\progmodrel, c, \mstate)
  \end{align*}
  then $\singletontrace{c} \continuedsim{\progmodrel} \singletontrace{\mstate}$ holds in the meta-logic.
\end{theorem}
In particular, we must show that the trace interpretation holds for the initial singleton traces, that the weakest precondition holds, and that the predicate $\alwaysholds(\progmodrel, c, \mstate)$ holds.
The predicate $\alwaysholds(\progmodrel, c, \mstate)$ is an \Iris predicate that states that $\xi$ (which is a relation on finite \emph{traces}, not mentioning resources of the logic) does in fact always hold, \emph{assuming} that all invariants hold, that postconditions hold (for any thread that has terminated up to that point in the trace), that the trace interpretation holds, and that the program never gets stuck.
We refer to our accompanying \Coq{} development for the precise definition of the $\alwaysholds$ predicate.
In practice, designing program logics on top of \Trillium{} is thus a matter of carefully picking a trace interpretation that admits user-friendly reasoning principles while allowing the user to conclude strong $\xi$ relations.\footnote{In fact, Trillium is conservative in that one can always pick a trivial model LTS, \raisebox{-0.2em}{\begin{tikzpicture}[scale=0.5, every node/.style={scale=0.5}] \draw node[circle, fill=black] (bullet) {}; \draw (bullet) edge[-stealth, in = 120, out = 70, looseness = 9] (bullet); \end{tikzpicture}}, and $\xi$ yielding no further intricacies in the program logic, and the regular safety guarantees of Iris.}

The adequacy theorem of \Trillium{} is much stronger than that of the ordinary \Iris{} program logic.
Recall that the usual adequacy theorem only establishes postcondition validity, progress, and invariant preservation.
\Trillium{}'s adequacy theorem, on the other hand, establishes intensional refinement for any relation $\xi$ which in turn can include all three aforementioned guarantees.
In addition, as per \Cref{lem:extending-int-ref}, it allows us to show that infinite traces of execution of the program also have corresponding infinite model traces.
Consequently, the adequacy theorem of \Trillium{} is much more complicated to establish.
The proof is in two stages.
We first show that the weakest precondition implies a finite approximation of intensional refinement stated in terms of guarded recursion of \Iris{}'s base logic.
We then show that the aforementioned guarded recursive definition implies intensional refinement---this is where we exploit relative image-finiteness.
We refer to the accompanying \Coq{} development for the details of the proof.

\section{The \Fairis logic}
\label{sec:fairis}

\newcommand{\fairrel}{\xi_{\mathit{fuel}}}
\newcommand{\tracefairrel}{\hat{\xi}_{\mathit{fuel}}}
\newcommand{\fairtermination}{\mathsf{fairly\_terminating}}
\newcommand{\FairlyTerminating}{\mathsf{FairlyTerminating}}
\newcommand{\fair}{\mathsf{fair}}
\newcommand{\first}{\mathsf{first}}
\newcommand{\terminates}{\mathsf{terminates}}
\newcommand{\initfuelres}{\mathsf{init\_fuel}}
\newcommand{\tofuel}{\mathsf{to\_fuel}}
\newcommand{\hasfuel}[2]{#1 \Mapsto #2}
\newcommand{\fmap}{\mathop{\$}}
\newcommand{\fuelsvar}{\mathit{fs}}
\newcommand{\prestepmod}{\upd^{\stateinterp}}
\newcommand{\finbranch}{\mathit{fin}}

In this section we present the \Fairis logic; an instantiation of Trillium for proving liveness properties under fairness assumptions, such as fair termination, of programs written in an OCaml-like programming language.
Most non-trivial concurrent programs only enjoy liveness properties, including termination, under fair thread scheduling.
To see this, consider the program in \Cref{fig:yes-no-program}.
In this example, the function~\lstinline|yn_start| creates a new reference~\lstinline|b| initialized
at~0, and starts two threads: \lstinline|yes| swaps the value of~\lstinline|b|
from $\True$ to $\False$ in a loop~\lstinline|k| times, and the \lstinline|no| swaps it from $\False$
to $\True$ as many times.
For thread \lstinline|yes| to make progress, it has to wait for \lstinline|no|
to set the value of~\lstinline|b| to~$\True$, and \emph{vice versa}.
Clearly, this program would not terminate with an unfair scheduler which only
gives execution time to, say, \lstinline|yes|.
But this program \emph{does} terminate in practice, because operating systems' schedulers and
processors do behave fairly.
We discuss the details of the proof of termination of the example in \Cref{fig:yes-no-program} in \Cref{sec:example_yesno}.

Our approach to taking fairness into account when reasoning about liveness properties is to incorporate it into the model LTS as follows.
First, we use the labels of the model LTS~$\model$, which we will henceforth call \emph{roles} in this section, in order to be able to express fairness at the level of the model.
Intuitively, a role is the abstraction of the concept of a thread at the model level.
Bear in mind, though, that, as we will discuss, threads and roles are not always in one-to-one correspondence.
To this end, we introduce a construction $\Fuel(\model)$ which augments the model~$\model$ to keep track of the correspondence between threads and roles in order to provide finite stuttering.
This construction also adapts the relation $\xi$ and lifts it to a relation $\Fuel(\xi)$ between traces of the program and traces of $\Fuel(\model)$.
For brevity, we will write $\fuelxi$ instead of $\Fuel(\xi)$.

In order to show that the program enjoys a certain liveness property $P$, \ie{}, that all \emph{fair} execution traces $\tau$ of the program satisfy $P(\tau)$, we carry out the proof in the following steps:
\begin{enumerate}
\item We choose a model~$\model$ and a relation~$\xi$.
\item We prove a specification for our program using the \Fairis{} program logic whose adequacy theorem ensures that all program traces (including unfair executions) refine some trace of the instrumented model~$\Fuel(\model)$ for the relation $\fuelxi$.
\item \label{itm:fairis-method-step2} We have proved (once and for all) that given two traces $\tau$ and $\tau'$, if $\fueltrhold{\tau}{\tau'}$, and $\tau$ is fair, then so is $\tau'$. (Intuitively, this holds because we have constructed $\Fuel(\model)$ and $\fuelxi$ such that they only allow finite stuttering in each thread.)
\item\label{itm:all-traces} We prove that all fair traces of $\model$ satisfy the desired property $P$.
\item\label{itm:P-preserved} We prove that~$P$ is preserved by~$\fuelxi$, in that $\fueltrhold{\tau}{\tau'}$ and $P(\tau')$ implies $P(\tau)$.
\item\label{itm:destutter} We prove that, if for a trace $\tau$ in $\Fuel(\model)$ we have that the corresponding trace $\tau'$ satisfies~$P$, then so does $\tau$, this roughly corresponds to the fact that $P$ is invariant under finite stuttering.
\item\label{itm:ccl} From all of the above, we conclude $P(\tau)$ for any fair trace $\tau$ of the program.
\end{enumerate}
This may seem like a lot of steps, but (\ref{itm:ccl}) is automatic, and given a fixed property~$P$, steps~(\ref{itm:P-preserved}) and~(\ref{itm:destutter}) can be proved independently of the program and of the model $\model$.
Step (\ref{itm:destutter}) is trivial for the case of termination and properties of the always-eventually form as these are preserved under finite stuttering.
Finally, in case of termination, we have formulated a simple criterion for step (\ref{itm:all-traces}) which is local in the sense that one does not need to consider traces but rather need to inspect individual transitions in~$\model$.
See our accompanying appendix and \Coq{} formalization for this local criterion.

\begin{figure}[t]
  \begin{minipage}[b]{.42\textwidth}
    \begin{AnerisPL}[numbers=none]
let rec yes b n =
  if cas b true false then n := !n-1;
  if !n > 0 then yes b n

let rec no b m =
  if cas b false true then m := !m-1;
  if !m > 0 then no b m

let yn_start k = let b = ref true in
  (yes b (ref k) || no b (ref k))
    \end{AnerisPL}
  \caption{The $\Yes$ and~$\No$ threads.}
  \label{fig:yes-no-program}
  \end{minipage}
  \vrule 
  \begin{minipage}[b]{.57\textwidth}
    \begin{center}
      \begin{tikzpicture}[every node/.style={scale=0.75}, scale=0.75]
        \draw (-3.7, -1.5) node[outer sep = 0.3em] (ldots) {$\mathbf{\dots}$};
        \draw (-1.7, -1.5) node[draw, circle, thick, inner sep = 0.2em, minimum width = 3em] (m_false) {\footnotesize $\mathbf{m, \bot}$};
        \draw (0, -1.5) node[draw, circle, thick, inner sep = 0.2em, minimum width = 3em] (m_true) {\footnotesize $\mathbf{m, \top}$};
        \draw (1.7, -1.5) node[draw, circle, thick, inner sep = 0.2em, minimum width = 3em] (m_plus_false) {\footnotesize $\mathbf{m\!+\!1, \bot}$};
        \draw (3.4, -1.5) node[draw, circle, thick, inner sep = 0.2em, minimum width = 3em] (m_plus_true) {\footnotesize $\mathbf{m\!+\!1, \top}$};
        \draw (5.1, -1.5) node[outer sep = 0.3em] (rdots) {$\mathbf{\dots}$};

        \draw (m_false) edge[-stealth, thick, dotted] node[sloped, above, scale=0.8] {$\mathsf{No}$} (ldots);

        \draw (m_false) edge[-stealth, thick, looseness = 4] node[sloped, above, scale=0.8] {$\mathsf{Yes}$} (m_false);
        \draw (m_true) edge[-stealth, thick] node[sloped, above, scale=0.8] {$\mathsf{Yes}$} (m_false);

        \draw (m_true) edge[-stealth, thick, looseness = 4] node[sloped, above, scale=0.8] {$\mathsf{No}$} (m_true);
        \draw (m_plus_false) edge[-stealth, thick] node[sloped, above, scale=0.8] {$\mathsf{No}$} (m_true);

        \draw (m_plus_false) edge[-stealth, thick, looseness = 4] node[sloped, above, scale=0.8] {$\mathsf{Yes}$} (m_plus_false);
        \draw (m_plus_true) edge[-stealth, thick] node[sloped, above, scale=0.8] {$\mathsf{Yes}$} (m_plus_false);

        \draw (m_plus_true) edge[-stealth, thick, looseness = 4] node[sloped, above, scale=0.8] {$\mathsf{No}$} (m_plus_true);

        \draw (rdots) edge[-stealth, thick, dotted] node[sloped, above, scale=0.8] {$\mathsf{No}$} (m_plus_true);

        \draw (0.5, -3.5) node {We write $\top$ for $\True$ and $\bot$ for $\False$.};
      \end{tikzpicture}
    \end{center}
  \caption{The model~$\genmod_{\mathsf{yes\_no}}$.}
  \label{fig:yes-no-model}
  \end{minipage}
\end{figure}


What the user obtains from the adequacy theorem of \Fairis{} is $\refines_{\fuelxi}$ between the program and the model LTS $\Fuel(\fmodel)$; $\fuelxi$ is relative image-finite by construction whenever $\xi$ is. Crucially, all the liveness properties we are interested in---such as termination or ``always-eventually'' trace properties---are stable under finite stuttering. Hence, if all fair traces of $\fmodel$ satisfy our property so do all fair traces of $\Fuel(\fmodel)$. Consequently, the user only has to prove that the property can be transported along the $\refines_{\fuelxi}$ intensional refinement relation from $\Fuel(\fmodel)$ to the program. In some cases, \eg{}, \cref{sec:example_evenodd}, we may be interested in proving a liveness property under an assumption stronger than fairness, \ie{}, that all fair traces that satisfy a certain property $Q$ satisfy the desired liveness property $P$. In such cases the user of \Fairis{} is responsible for proving that the property $Q$ can be transported from the program to $\Fuel(\fmodel)$ and from $\Fuel(\fmodel)$ to $\fmodel$---in case of the example in \cref{sec:example_evenodd}, we are interested in infinite fair traces; infinitude, like finiteness, trivially transports as it is preserved by intensional refinement and finite stuttering. The $\Fuel$ construction essentially augments the model LTS to add two pieces of information: (1) for each thread id~$\zeta$, the set of roles that are associated to $\zeta$, and, (2) for each role~$\rho$, the value of its so-called \emph{fuel}, \ie{}, a number which measures how long the program can still postpone taking a step in that role in $\xi$. 
%
This allows us to assign \emph{obligations} to threads, \ie{}, assign a set of
\emph{live} roles to a thread---a role is said to be live if it can still take
steps. A thread that has a set $R$ of roles assigned to it must eventually
(restricted by fuel) take a step in each of those roles. This is enforced by
mandating that to take a step of computation, a thread must have a non-empty
set~$R$ of live roles, and that after the step all fuels in~$R$ are decreased,
except if this step was matched by a non-stuttering step corresponding to a role
$\rho \in R$ (one of its live roles) in the underlying LTS $\fmodel$. In that
case, the thread may increase the fuel of role $\rho$ (but the fuel for all other roles in $R$
must decrease) as long as it remains under a certain global cap which we will
denote with $\fuellimit$.%
\footnote{This global cap, fixed for each refinement
proof, ensures relative image-finiteness (\cref{def:rel-img-fin}) of
$\Fuel(\fmodel)$.}
Alternatively, a thread can \emph{delegate} some of its obligations to a thread
that it forks. The adequacy theorem of \Fairis{} requires a proof in the program
logic that enforces that threads may only stop if there are no live roles
assigned to them anymore; this is encoded through threads' postconditions.


%
The \Fairis adequacy theorem is a special case of the Trillium adequacy theorem.
The theorem is shown below, and has differences from the Trillium adequacy
theorem {\wphighlight highlighted}:
\begin{theorem}[Fairis-Adequacy]\label{thm:fairis_adequacy}
  Let $\expr$ be a program, $\progstate$ a program state, $\zeta$ the locale of $\expr$ in an otherwise empty thread pool.
  Let $\mstate \in {\wphighlight\fmodel}$ be a model state, and $\progmodrel$ a relative image-finite relation on program traces and model traces.
  Let $c = (\set{\zeta \mapsto \expr}, \progstate)$ be the initial configuration of the program, {\wphighlight and $\initialfuelstate{\mstate}$ be the initial state of $\Fuel(\fmodel)$ corresponding to $\mstate$, \ie{}, $\mstate$ together with
    $\zeta$ assigned to a map $\fuellimitmap$ of all roles with
    $\fuellimit$ fuel}.
  If $\xi(\singletontrace{c}, \singletontrace{\mstate})$ and
  \begin{align*}
    \pvs[\top]
    \begin{aligned}[t]
      & 
      {\wphighlight
        (\fairmodelfrag(\mstate) \ast \hasfuel{\zeta}{\fuellimitmap} \ast
      \SSep_{\loc;\val \in \sigma} \loc \mapsto \val) \wand }\;\\
      &
      \stateinterp(\singletontrace{c}, \singletontrace{\initialfuelstatehighlight{\mstate}}) \ast
      \twpre{\expr}[\top]{{\wphighlight \Ret \valB. \hasfuel{\zeta}{\emptyset}}}[\zeta]
      \ast \alwaysholds(\progmodrel, c, \initialfuelstatehighlight{\mstate})
    \end{aligned}
  \end{align*}
  then $\singletontrace{c}
  \continuedsim{\wphighlight\fuelxi}
  \singletontrace{{\initialfuelstatehighlight{\mstate}}}$ holds in the meta-logic.
\end{theorem}
Here, $\fairmodelfrag(\mstate)$ and $\hasfuel{\tid}{\fuellimitmap}$ are initial resources used in the \Fairis program logic rules.
In particular, $\fairmodelfrag(\mstate)$ is an exclusive resource that tracks
the current state of the user model, while $\hasfuel{\zeta}{\fuelsvar}$
associates the locale $\zeta$ with a partial map $\fuelsvar : \mathsf{Role(\genmod)} \rightharpoonup \nat$ that associates model roles
with their fuels.
The authoritative part $\fairmodelfull(\mstate)$ of the resource is kept as part of the trace interpretation; the name $\gname_\fmodel$ is a globally fixed name for the entire proof created as part of the proof of the adequacy theorem.
The resource $\fairmodelfrag(\mstate)$ is available to the user of the \Fairis{} program logic to be able to relate the state of the program to the state of the model (which always agrees with the one tracked in the trace interpretation as per resource rules), usually in an invariant.
This relation in the logic is then what the user will use to establish the $\xi$ relation as required---we will see this in examples below.
Initially, all roles are assigned to the single thread in the thread-pool with fuels all being $\fuellimit$.
The last piece of resource the user acquires from the adequacy theorem is the
full ownership of the entire initial state (heap) $\sigma$ in that we obtain points-to propositions for all locations in $\sigma$.
The adequacy theorem, and the rule for forking threads as we will see later, enforce that upon termination (in the postcondition) the thread must own no live roles.
Finally, note that even though the adequacy theorem establishes a
$\continuedsim{\wphighlight\fuelxi}$ relation, the user only needs to prove
$\alwaysholds$ for $\xi$ and not for $\fuelxi$.

In the rest of this section we cover how we apply \Fairis to prove various liveness properties.
First, we present the key ideas of the \Fairis program logic, along with its reasoning rules, used for proving the weakest precondition of programs \Cref{sec:fairis_program_logic}.
We then give a tour of how the \Fairis program logic can be used to prove liveness properties of programs \Cref{sec:fairis_examples}, including fair termination of the yes-no example discussed earlier.
Finally, we outline some of the technical details involved in obtaining the
Fairis adequacy theorem \Cref{sec:fairis_adequacy}.

The examples that we present in this section are simple examples designed to demonstrate viability of the Trillium approach to liveness properties under fair scheduling assumptions.
As such, in all these examples the LTS model used includes the entire core functionality of the program; only the administrative program steps, \eg{}, beta-reduction, are not included which are taken into account via stuttering enabled by the fuel construction.
Hence, in that sense, the LTS models of these examples are not very abstract.
For more involved examples, we expect that the level of abstraction of the LTS model compared to the program should be similar to that of \Aneris{} examples; c.f. the Paxos example of \cref{sec:paxos-clients} where the model abstracts away a substantial amount of implementation details.
However, to substantiate this claim one needs a proper experimental evaluation, using a wide range of more advanced examples, which is beyond the scope of the current work, and thus we leave it for future work.

\subsection{The \Fairis Program Logic}
\label{sec:fairis_program_logic}

\newcommand{\successor}{+1}
\newcommand{\fmapincr}[1]{#1^{++}}

The \Fairis program logic combines the reasoning principles of conventional Iris program logics with reasoning principles involving fuel and model resources.
To properly compartmentalize these orthogonal reasoning principles, the program logic employs two layers of reasoning: an outer (model) layer and an inner (program) layer.

The intuition is that at every program step, the program will either do a \emph{fuel step}, corresponding to stuttering, or a \emph{model step}, corresponding to a step in the underlying model.
Both of these are handled by the outer model layer.
Regardless of which kind of step the program takes, the inner layer will then be used to reason about the actual program step.
The rules of the inner program logic only concern the program on its own and hence very closely resemble those of the ordinary program logic of \Iris{}.

The inner program logic layer is expressed in terms of a different weakest precondition
$\sstwpre{\expr}[\mask]{\pred}$ (note the angle brackets instead of braces for the postcondition).
This weakest precondition is a (heavily) simplified version of \Trillium{}'s weakest precondition where the state interpretation (it indeed only mentions the state and not the entire trace) only concerns the heap of the program.
In particular, this inner program logic on its own would be useless.
It is only useful, and extremely so, because the outer program logic has already taken care of the details of the model including the fuel.
Additionally, the inner weakest precondition strictly captures a \emph{single} step of reduction as for further steps the outer program logic must intervene again to manage the model-side details.
As a result, the postcondition of the inner weakest preconditions are predicates over arbitrary expressions as opposed to values, \ie{}, a predicate $\pred : \Expr \to \iProp$.
Similar to the outer weakest precondition, we often write
$\sstwpre{\expr}[\mask]{\Ret \expr'. \propB} \triangleq
\sstwpre{\expr}[\mask]{\Lam \expr'. \propB}$, and
$\sstwpre{\expr}[\mask]{\propB} \triangleq
\sstwpre{\expr}[\mask]{\Ret \expr'. \expr' = \TT \ast \propB}$.
Additionally, we often use $\valB$ instead of $\expr'$ for the return value
binder, to denote that the returned expression is a value.
The definition of the inner weakest precondition is given in
the accompanying appendix.

\begin{figure}
  \textbf{Fuel \& model rules (outer program logic)}
  \begin{mathpar}
    \inferH{wp-step-fuel}
    { { \begin{array}{@{}c@{}}
      \hasfuel{\zeta}{\fmapincr{\fuelsvar}} \and \fuelsvar \neq \emptyset \\
      \sstwpre{\expr}[\mask]{\Ret \expr'. \hasfuel{\zeta}{\fuelsvar} \wand
        \twpre{\expr'}[\mask]{\pred}[\zeta]}
    \end{array}}}
    {\twpre{\expr}[\mask]{\pred}[\zeta]}
    \and
    \inferH{wp-role-dealloc}
    { {\begin{array}{@{}c@{}}
          \fairmodelfrag(\mstate) \and
        \mstate \lto{\rho}\!\!\,\!\!\!\!\!\scalebox{.7}{/} \, \ \ \_ \and
      \hasfuel{\zeta}{\lbrace\rho := \_\rbrace \uplus \fuelsvar}
      \\
      (\fairmodelfrag(\mstate) \ast
      \hasfuel{\zeta}{\fuelsvar}) \wand
      \twpre{\expr}[\mask]{\pred}[\zeta]
      \end{array}}}
    {\twpre{\expr}[\mask]{\pred}[\zeta]}
    \and
    \inferH{wp-step-model}
    { {\begin{array}{@{}c@{}}
      \fairmodelfrag(\mstate) \and
     \mstate \lto{\rho} \mstate' \and
        \hasfuel{\zeta}{\lbrace\rho := \_\rbrace \uplus
          (\fmapincr{\fuelsvar})}
        \\
        \sstwpre{\expr}[\mask]{
        \Ret \expr'.
        (\fairmodelfrag(\mstate') \ast
        \hasfuel{\zeta}{\lbrace\rho := \fuellimit\rbrace \uplus \fuelsvar}) \wand
        \twpre{\expr'}[\mask]{\pred}[\zeta]}
    \end{array}}}
    { \twpre{\expr}[\mask]{\pred}[\zeta]}
    \quad
    \inferH{wp-role-fork}
    { {\begin{array}{@{}c@{}}
      \hasfuel{\zeta}{\fmapincr{(\fuelsvar_1 \uplus \fuelsvar_2)}} \and
      \fuelsvar_1 \uplus \fuelsvar_2 \neq \emptyset \\
      \All \zeta'.
      \hasfuel{\zeta'}{\fuelsvar_2} \wand
      \twpre{\expr}[\mask]{\hasfuel{\zeta'}{\emptyset}}[\zeta']
    \end{array}}}
    {\twpre{\Fork{\expr}}[\mask]
      {\hasfuel{\zeta}{\fuelsvar_1}}[\zeta]}
  \end{mathpar}
  \textbf{Program rules (inner program logic; an excerpt)}
  \begin{mathpar}
  \axiomH{wp-alloc}
  {\sstwpre{\Alloc\ \val}[\mask]{\valB. \Exists \loc. \valB = \loc \ast \mapsto \val}}
  \and
  \inferH{wp-load}
  {\loc \mapsto \val}
  {\sstwpre{\deref \loc}[\mask]{\Ret \valB. \valB = \val \ast \loc \mapsto \val}}
  \and
  \inferH{wp-store}{\loc \mapsto \val}{\sstwpre{\loc \gets \valB}[\mask]{\loc \mapsto \valB}}
  \and
  \inferH{wp-cas}
  {\loc \mapsto \val}
  {\sstwpre{\Cas{\loc}{\val_1}{\val_2}}[\mask]{
      \begin{aligned}
        \Ret \valB.
        \begin{aligned}[t]
          &(\val = \val_1 \ast \valB = \True \ast \loc \mapsto \val_2)\ \vee\\[-0.3em]
          &(\val \neq \val_1 \ast \valB = \False \ast \loc \mapsto \val)
        \end{aligned}
  \end{aligned}}}
  \and
  \inferH{wp-beta}{}{\sstwpre{(\lambda \var.\; \expr)~\val}[\mask] {\Ret \expr'. \expr' = \expr[\val/\var]}}
\end{mathpar}
\vspace{-1.5em}
  \caption{The rules of the \Fairis program logic}
  \label{fig:fairis}
\end{figure}

An overview of the layers and their rules can be found in \Cref{fig:fairis}.
The rules of the inner program logic are entirely standard.
We will thus not explain them in detail.
The \ruleref{wp-step-fuel} rule captures that we can take a \emph{fuel step} whenever the fuel of all the roles of the non-empty fuel map is non-zero ($\hasfuel{\zeta}{\fmapincr{\fuelsvar}}$ and $\fuelsvar \neq \emptyset$, where $\fmapincr{\fuelsvar}$ denotes the map $\fuelsvar$ where all fuels are incremented by 1).
We must then prove the inner weakest precondition, where the postcondition captures that we re-obtain the fuel map ($\hasfuel{\zeta}{\fuelsvar}$), where all fuel has been decremented, for the remaining proof obligation.

The \ruleref{wp-step-model} rule captures that we can take a \emph{model step} whenever the underlying model can take a step over some role $\rho$; we have $\fairmodelfrag(\mstate)$ indicating the current model state, and $\mstate \lto{\rho} \mstate'$ indicating we can take a step.
We must then show that the role $\rho$ is associated with the current locale $\zeta$, and that the fuel of all non-$\rho$ roles are non-zero ($\hasfuel{\zeta}{\lbrace\rho := \_\rbrace \uplus \fmapincr{\fuelsvar}}$).
Just as in \ruleref{wp-step-fuel} we must show the inner weakest precondition, which in turn gives us back the model and fuel resources.
The model resource is updated to the new model state, $\fairmodelfrag(\mstate')$, and the fuel map has the fuel of $\rho$ replenished to the maximum fuel cap, $\fuellimit$, while all non-$\rho$ fuels have been decremented, $\hasfuel{\zeta}{\lbrace\rho := \fuellimit\rbrace \uplus \fuelsvar}$.
The conjunction of these rules effectively enforce, that for any role $\rho$ for which we can only take finitely many steps in the thread that $\rho$ is assigned to, we are forced to take a step corresponding to $\rho$.
Otherwise, we would run out of fuel for the role $\rho$ and hence not be able to take anymore steps, be it a stuttering step, or a step of any role other than $\rho$.

The rules \ruleref{wp-role-dealloc} and \ruleref{wp-role-fork} allow us to discharge our obligations (roles we are responsible for).
The rule \ruleref{wp-role-dealloc} removes a role from the fuel map of the locale when all the obligations of that role are fulfilled, \ie{}, when the role ``terminates'' and cannot take any more steps.
The rule \ruleref{wp-role-fork}, on the other hand, allows the thread to \emph{delegate} some of its obligations (roles) by forking a thread and passing those roles to the newly forked threads.
Note how the fuel is decremented for all roles, including those that are passed to the newly forked threads.
This restriction is necessary for the soundness of the logic as otherwise one could indefinitely postpone taking a step in role by repeatedly forking new threads.
Furthermore, note how the rule \ruleref{wp-role-fork} requires the postcondition of the forked thread to have a live role when it terminates, effectively forcing it to discharge all its obligations before termination.








\subsection{Examples}
\label{sec:fairis_examples}

In this section we cover a suite of examples to demonstrate how \Fairis can be
used to prove fairness-dependent liveness properties.
In particular, we first prove \emph{fair termination} of the yes-no example (\Cref{fig:yes-no-program}) discussed earlier.
We will then present an example in \Cref{sec:example_choosenat} where the model LTS is \emph{not} image-finite but still the $\xi$ relation is relative image-finite \cf{} \Cref{def:rel-img-fin}.
Finally, we demonstrate how \Fairis can be used to show liveness properties beyond termination in \Cref{sec:example_evenodd}.

\subsubsection{Fair Termination: Yes-No}
\label{sec:example_yesno}

\newcommand{\yesname}{\gname_{\mathit{yes}}}
\newcommand{\noname}{\gname_{\mathit{no}}}
\newcommand{\flagloc}{l}
\newcommand{\yesrole}{\mathsf{Yes}}
\newcommand{\norole}{\mathsf{No}}

We initially consider the application of \Fairis for proving fair termination,
which we formally state as follows over the initial configuration $c$ of the program:
\begin{align*}
  \fairtermination(c) \triangleq \All \tau. \first(\tau) = c \implies \fair(\tau) \implies \finite(\tau)
\end{align*}
In particular, we prove fair termination for the yes-no program presented in \Cref{fig:yes-no-program}, using the associated model shown in \Cref{fig:yes-no-model}.
Proving fair termination of the model is relatively straightforward.
In our \Coq{} formalization we apply the local criterion mentioned earlier; this criterion is explained in our accompanying appendix.
Moreover, as discussed earlier, termination can be trivially transported from $\fmodel$ to $\Fuel(\fmodel)$.
Hence, to prove fair termination of the yes-no program, we just have to show that it is refined by the yes-no model using \Fairis{}.

To do so, we apply the adequacy theorem of \Fairis, using a trivial $\xi$ relation, and an empty initial state $\sigma$ for the program.
This choice means that $\alwaysholds(\Fuel(\xi),c,\initialfuelstate{\mstate})$ holds trivially for any $c$ and $\initialfuelstate{\mstate}$.
Since the yes-no model (\Cref{fig:yes-no-model}) is finitely branching, any $\xi$ relation on it is trivially relative image-finite.
What remains to prove is the weakest precondition of the yes-no program:
\[
(\fairmodelfrag(\mstate) \ast \hasfuel{\zeta}
  {\lbrace\yesrole := \fuellimit; \norole := \fuellimit\rbrace}) \wand
\twpre{\mathsf{yn\_start}\ n}[\top]{\hasfuel{\zeta}{\emptyset}}[\zeta]
\]
where $m$ is the state $(n, \top)$ in \Cref{fig:yes-no-model}, and $0 < n$ is any \emph{positive} natural number.
As for applying the rules updating the underlying model, the only steps in the program where this happens are the two $\langkw{cas}$ operations in the yes and no threads.
These are precisely the horizontal steps in \Cref{fig:yes-no-model} when the $\langkw{cas}$ operation succeeds and the loop when the $\langkw{cas}$ operation fails---as $\langkw{cas}$ is \emph{atomic}, we can access the invariant (see the rest of the proof argument) during a $\langkw{cas}$ operation.
All the other steps in the program are stuttering steps.

The proof starts by using \ruleref{wp-alloc} to resolve the allocation of the flag reference, obtaining $\flagloc \mapsto \True$.
As per conventional Iris methodology for concurrent programs, the proof employs an invariant to safely share the flag reference among the two threads.
This invariant additionally ensures that the program behaves according to the model LTS by incorporating the $\fairmodelfrag$ part of the model resource.
Recall that $n$ corresponds to how many times the model must cycle between flipping the flag $b$ back and forth before termination.
In order to tie the state of the model to the references managed by the two threads, we incorporate two additional resource predicates $\authfull_{\yesname}$ and $\authfull_{\noname}$ in the invariant, each tracking how many times each thread must still flip the flag $b$---the fragmental parts of these resources will be passed to the two threads.
The invariant used to prove the yes-no program, $\mathsf{yesno\_inv}(\yesname,\noname,\flagloc)$, is as follows:
\begin{align*}
  \mathsf{yesno\_inv}(\yesname,\noname,\flagloc) \triangleq{}
  & \Exists n, b. \fairmodelfrag{(n,b)} \ast (n,b) \neq (0,\False) \ast \flagloc \mapsto b\ \ast \\
  & \left(b = \True \ast \authfull_{\yesname}(n) \ast \authfull_{\noname}(n)\right) \lor \left(b = \False \ast \authfull_{\yesname}(n) \ast \authfull_{\noname}(n-1)\right)
\end{align*}
Note how in the case the flag is $\False$ the value of the no thread may lag behind because it has not yet had the chance to flip the flag in this ``round'' to be able to catch up.
Furthermore, note how this invariant asserts that the state $(0, \False)$ is not ever reached in the model when the program is refining the model.
This is crucially the case because we only ever call $\mathsf{yn\_start}$ with a strictly positive argument.
It only remains to prove the following specifications of the two threads
\[
\begin{array}[t]{@{}r@{\ }l@{}}
(\knowInv{\namesp_{\mathit{YN}}}{\mathsf{yesno\_inv}(\yesname,\noname,\flagloc)} \ast
\hasfuel{\zeta}{\lbrace \yesrole := 42 \rbrace} \ast
\authfrag_{\yesname}(n) \ast
0 < n) \wand &
\twpre{\mathsf{yes}\ \flagloc\ n}[\top]
{\hasfuel{\zeta}{\emptyset}}[\zeta]\\
(\knowInv{\namesp_{\mathit{YN}}}{\mathsf{yesno\_inv}(\yesname,\noname,\flagloc)} \ast
\hasfuel{\zeta}{\lbrace \norole := 42 \rbrace} \ast
\authfrag_{\noname}(n) \ast
0 < n) \wand &
\twpre{\mathsf{no}\ \flagloc\ n}[\top]
{\hasfuel{\zeta}{\emptyset}}[\zeta]
\end{array}
\]
Note that the fuel of $42$ is picked somewhat arbitrarily; it only needs to be below $\fuellimit$ (which we can arbitrarily pick) and large enough for the proof to go through.

We use these specifications along with \ruleref{wp-role-fork} to delegate the role obligations to the two threads, which completes the proof of the main thread which can (and must) now terminate as it has no roles associated to it anymore.
We use \ruleref{wp-step-fuel} to resolve all the stuttering steps in between model steps.
The crux of the rest of the proof is the $\langkw{cas}$ operations, where we update the model using the rule \ruleref{wp-step-model}, regardless of it succeeding or not.
In particular, we open the invariant around the operation using the rule \ruleref{inv-access}, and then resolve the operation with \ruleref{wp-cas}.
Based on whether the $\langkw{cas}$ operation succeeds, we either take the decrementing or the looping step in the model.
In either case, the invariant is preserved.
If we succeed, we update our model resource $\authfrag_{\gname_\fmodel}(n)$ accordingly to preserve the invariant.
If we fail, we simply do nothing and loop.
In both cases, a model step has taken place in the underlying logic and hence the fuel for thread's role can be replenished.
The program loops until we eventually hit 0, in which case we can use \ruleref{wp-role-dealloc} to discharge the role obligation, as the role can no longer step in the model, thus fulfilling the postcondition.

\subsubsection{Sound Infinite Branching: Non-deterministic Nat}
\label{sec:example_choosenat}

\newcommand{\nondetrole}{\mathsf{nondet}}

\begin{figure}[t]
  \begin{minipage}[b]{.5\textwidth}
    \begin{AnerisPL}[numbers=none]
let rec decr_loop l =
  if !l > 0 then l := !l - 1 else #()

let nondet_start l =
  l := (nondet () + 1);; decr_loop l
\end{AnerisPL}
\vspace{-1em}
  \caption{The non-deterministic nat example.}
  \label{fig:nondet_program}
  \end{minipage}
  \vrule
  \begin{minipage}[b]{.47\textwidth}
    \begin{center}
      \begin{tikzpicture}[every node/.style={scale=0.7}, scale=0.7]
        \draw (0, 0) node[draw, circle, thick, inner sep = 0.2em] (inf) {$\mathbf{\infty}$};
        \draw (-3, -1.5) node[draw, circle, thick, inner sep = 0.2em] (zero) {$\mathbf{0}$};
        \draw (-1.5, -1.5) node[draw, circle, thick, inner sep = 0.2em] (one) {$\mathbf{1}$};
        \draw (0, -1.5) node[draw, circle, thick, inner sep = 0.2em] (two) {$\mathbf{2}$};
        \draw (1.5, -1.5) node[draw, circle, thick, inner sep = 0.2em] (three) {$\mathbf{3}$};
        \draw (3.3, -1.5) node[outer sep = 0.3em] (dots) {$\mathbf{\dots}$};

        \draw (inf) edge[-stealth, thick] node[sloped, above, scale=0.6] {$\mathsf{nondet}$} (zero);

        \draw (inf) edge[-stealth, thick] node[sloped, above, scale=0.6] {$\mathsf{nondet}$} (one);
        \draw (one) edge[-stealth, thick] node[sloped, above, scale=0.6, shift={(.5em, 0)}] {$\mathsf{nondet}$} (zero);

        \draw (inf) edge[-stealth, thick] node[sloped, above, scale=0.6] {$\mathsf{nondet}$} (two);
        \draw (two) edge[-stealth, thick] node[sloped, above, scale=0.6, shift={(.5em, 0)}] {$\mathsf{nondet}$} (one);

        \draw (inf) edge[-stealth, thick] node[sloped, above, scale=0.6] {$\mathsf{nondet}$} (three);
        \draw (three) edge[-stealth, thick] node[sloped, above, scale=0.6, shift={(.5em, 0)}] {$\mathsf{nondet}$} (two);

        \draw (inf) edge[-stealth, thick, dotted] node[sloped, above, scale=0.6] {$\mathsf{nondet}$} (dots);
        \draw (dots) edge[-stealth, thick, dotted] node[sloped, above, scale=0.6] {$\mathsf{nondet}$} (three);
      \end{tikzpicture}
    \end{center}
    \vspace{-1em}
    \caption{The non-deterministic nat model.}
  \label{fig:nondet_model}
  \end{minipage}
\end{figure}

In \Cref{sec:introduction} we remarked that the model should be relatively image-finite for the Trillium (and thereby \Fairis) adequacy theorem to be sound.
In this section we consider a model which is infinitely branching, while still being relative image-finite, by virtue of the user picked relation $\xi$.

The program we consider can be seen in \Cref{fig:nondet_program}.
This program works on a pre-allocated reference $\loc$ as input.
It then assigns a non-zero non-deterministically chosen natural number
$n$ to the reference, which it decrements until it hits zero,
after which point it terminates.
A specification for the non-deterministic number operation is simply $\sstwpre{\mathsf{nondet}}[\mask]{\valB. \Exists n. \valB = n}$.
The model of the program can be seen in \Cref{fig:nondet_model}.
It starts in an initial state $\infty$, from which it can go to any natural number $n$.
While the model is terminating (all traces are finite), it is also infinitely branching to accommodate for the fact that the program may non-deterministically step to any natural number.

In order to prove relative image-finiteness we apply the \Fairis adequacy theorem
with the following user relation:
\begin{align*}
  \xi_{\mathsf{nondet}}^{\loc}(\tau, \kappa) \triangleq
  (\TraceLast(\kappa) = \infty \wedge \mathsf{heap}(\TraceLast(\tau))(\loc) = -1) \vee (\Exists n. \TraceLast(\kappa) = n \wedge \mathsf{heap}(\TraceLast(\tau))(\loc) = n)
\end{align*}
That is, the state of the model always correspond to the number stored in the location, once it has been initialized---before that the location stores $-1$.
This relation ensure relative image-finiteness, as for every program step, there is only \emph{one} valid model step.
Intuitively, the model side number we step to is uniquely (and hence finitely) determined by the program.

To be able to prove
$\alwaysholds(\xi_\mathsf{nondet}(\loc), c, \mstate)$
we allocate a resource predicate, starting in the initial model state
$\authfull_{\gname}(\infty)$ and $\authfrag_\gname(\infty)$, and use the former establish the following invariant at the very beginning of our verification:
\[
\mathsf{nondet\_inv}(\gname,\loc) \triangleq
\Exists \mathit{cn}.
\fairmodelfrag(\mathit{cn}) \ast \authfull_{\gname}(\mathit{cn}) \ast
(\mathit{cn} = \infty \ast \loc \mapsto -1) \vee
(\Exists n.\; \mathit{cn} = n \ast \loc \mapsto n)
\]
This invariant enforces that the location points to
a natural number corresponding to the model.
Additionally, whenever the model is in the initial $\infty$ state, the
location stores $-1$.
Note that the relation $\xi$ follows directly from the invariant.

To prove fair termination of the program, it only remains to prove the following specification:
\[
(\knowInv{\namesp_{\mathsf{nondet}}}{\mathsf{nondet\_inv}(\gname,\loc)} \ast
\authfrag_{\gname}(\infty) \ast
\hasfuel{\zeta}{\lbrace \nondetrole := \fuellimit\rbrace}) \wand
\twpre{\mathsf{nondet\_start}\ \loc}[\top]{\hasfuel{\zeta}{\emptyset}}[\zeta]
\]
It is straightforward to prove that the invariant holds throughout the program,
and that it implies the user defined relation.
The only non-stuttering step in the proof is the assignment step in the then branch of \lstinline|decr_loop| which allows us to replenish the fuel in every loop iteration.
An interesting aspect of this proof is that we use an invariant in the proof despite the fact that program is not concurrent.
In this case, the invariant is not used to facilitate sharing between threads but rather to enforce \emph{an invariant} of the program, \ie{}, that the value of the reference must always, after initialization, correspond to what is determined in the model LTS.

\subsubsection{Liveness Properties Beyond Termination: Even-Odd}
\label{sec:example_evenodd}

\begin{figure}[t]
  \begin{minipage}[b]{.64\textwidth}
    \begin{AnerisPL}[numbers=none]
let rec incr_loop l n =
  if cas l n (n+1) then incr_loop l (n+2) else incr_loop l n

let eo_start l =
  let x := !l in (incr_loop l x || incr_loop l (x+1))
\end{AnerisPL}
\vspace{-1em}
  \caption{The even-odd example.}
  \label{fig:evenodd_program}
  \end{minipage}
  \vrule
  \begin{minipage}[b]{.35\textwidth}
    \begin{center}
      \begin{tikzpicture}[every node/.style={scale=0.65}, scale=0.65]
        \draw (-3, -1.5) node[draw, circle, thick, inner sep = 0.2em] (zero) {$\mathbf{0}$};
        \draw (-1.5, -1.5) node[draw, circle, thick, inner sep = 0.2em] (one) {$\mathbf{1}$};
        \draw (0, -1.5) node[draw, circle, thick, inner sep = 0.2em] (two) {$\mathbf{2}$};
        \draw (1.5, -1.5) node[draw, circle, thick, inner sep = 0.2em] (three) {$\mathbf{3}$};
        \draw (3.3, -1.5) node[outer sep = 0.3em] (dots) {$\mathbf{\dots}$};

        \draw (zero) edge[-stealth, thick] node[sloped, above, scale=0.6] {$\mathsf{even}$} (one);
        \draw (zero) edge[-stealth, thick, looseness = 6] node[sloped, above, scale=0.6] {$\mathsf{odd}$} (zero);

        \draw (one) edge[-stealth, thick] node[sloped, above, scale=0.6] {$\mathsf{odd}$} (two);
        \draw (one) edge[-stealth, thick, looseness = 6] node[sloped, above, scale=0.6] {$\mathsf{even}$} (one);

        \draw (two) edge[-stealth, thick] node[sloped, above, scale=0.6] {$\mathsf{even}$} (three);
        \draw (two) edge[-stealth, thick, looseness = 6] node[sloped, above, scale=0.6] {$\mathsf{odd}$} (two);

        \draw (three) edge[-stealth, thick, dotted] node[sloped, above, scale=0.6] {$\mathsf{odd}$} (dots);
        \draw (three) edge[-stealth, thick, looseness = 6] node[sloped, above, scale=0.6] {$\mathsf{even}$} (three);
      \end{tikzpicture}
    \end{center}
  \caption{The even-odd model.}
  \label{fig:evenodd_model}
  \end{minipage}
\end{figure}

\newcommand{\fairprop}{\mathsf{fair\_prop}}
\newcommand{\evenoddprop}{\mathsf{evenodd\_prop}}
\newcommand{\evenoddpropmdl}{\mathsf{evenodd\_prop\_mdl}}
\newcommand{\maximal}{\mathsf{maximal}}
\newcommand{\evenname}{\gname_{\mathit{e}}}
\newcommand{\oddname}{\gname_{\mathit{o}}}

In this section we demonstrate how \Fairis{} can be used to prove liveness properties beyond termination. 

Consider the program shown in \Cref{fig:evenodd_program}.
The program works on a location $\loc$ assumed to store 0 initially.
It forks off two threads that each indefinitely increment the location whenever it is even or odd, respectively.
The property that we prove is that the counter $\loc$ visits \emph{all} natural numbers \emph{in order}.
This property only makes sense for maximal traces, \ie{}, traces that are either infinite or can take no further step---as we will discuss, all maximal traces of this program are infinite.
Formally, the property we prove is the following:
\begin{align*}
  \evenoddprop(\loc,\conf) & \triangleq{} \All \extr. \first(\extr) = \conf \Ra \fair(\extr) \Ra \maximal(\extr) \Ra\\
  & (\All i. \Exists n. \mathsf{heap}(\extr(n))(\ell) = i)\ \wedge & \textit{(all numbers visited)}\\
  & (\All n. \Exists i. \mathsf{heap}(\extr(n))(\ell) = i \wedge \Exists j. i \le j \wedge \extr(n+1)(\ell) = j) & \textit{(monotonicity)}
\end{align*}
We prove the property above by relating the program in \Cref{fig:evenodd_program} with the model LTS \Cref{fig:evenodd_model}.
To this end, we prove the following property about the model \Cref{fig:evenodd_model}.
\begin{align*}
  \evenoddpropmdl(\loc,\mstate) & \triangleq{} \All \kappa. \first(\kappa) = \mstate \Ra \fair(\kappa) \Ra \mathsf{infinite}(\kappa) \Ra\\
  & (\All i. \Exists n. \extr(n) = i)\ \wedge & \textit{(all number visited)}\\
  & (\All n. \Exists i. \extr(n) = i \wedge \Exists j. i \le j \wedge \extr(n+1) = j) & \textit{(monotonicity)}
\end{align*}
First, the property $\evenoddpropmdl(\loc,\mstate)$ is stable under finite stuttering---monotonicity is expressed using $\le$ which is preserved by stuttering.
Second, notice the discrepancy between this property and $\evenoddprop(\loc,\conf)$ where the latter only assumes maximality of the trace while the former requires infinitude.
This property is easy to prove under fairness assumptions.
A fair infinite trace cannot have a tail that just consists with one of the loops.
Hence, it must visit all numbers.
Monotonicity follows rather trivially.

To finish the proof, we need to establish that (1) any maximal execution trace is infinite, and (2) that the liveness property can be transported from the fuel-instrumented model to the execution trace.
To this end, we will pick $\xi$ relations as follows:
\begin{align*}
  \xi_{\mathsf{even\_odd}}^{\loc}(\tau, \kappa) \triangleq{} & \xi_{\mathsf{steps}}(\extr) \wedge \xi_{\mathsf{match}}^{\loc}(\extr,\kappa)\\
  \xi_{\mathsf{steps}}(\tau) \triangleq{} & \exists \zeta, \conf'.\; \tracelast(\tau) \lto{\zeta} \conf'\\
  \xi_{\mathsf{match}}^{\loc}(\extr, \kappa) \triangleq{} & \All n < \length(\extr). \extr(n)(\loc) = \kappa(n)
\end{align*}
Note that, again, because the model in \Cref{fig:evenodd_model} is finitely branching, the relation $\xi_{\mathsf{even\_odd}}$ is trivially relative image-finite.
Moreover, if we have $\tau~\tracepredof{\refines}_{\xi_{\mathsf{even\_odd}}^{\loc}}~\kappa$, given the $\xi_{\mathsf{steps}}(\tau)$ part of $\xi_{\mathsf{even\_odd}}^{\loc}$ together with maximality of $\tau$, we can conclude that $\tau$ must be infinite.
This established by (1) above.
The other part of the relation we pick, $\xi_{\mathsf{match}}^{\loc}$, simply expresses that the value stored in reference $\loc$ is always the same as the number in the model trace.
In order to prove $\alwaysholds(\xi_{\mathsf{even\_odd}}^{\loc},c,\mstate)$ we must show that $\xi_{\mathsf{steps}}^{\loc}$ is preserved throughout program execution.
This, however, is a simple consequence of post-condition validity and progress properties that we can assume as we prove $\alwaysholds$.
On the one hand, we know that the program cannot be stuck, that is each thread is either a value or it can take further steps.
On the other hand, we cannot have that all threads are values, as if they were, then their postconditions would all hold.
This would allow us to conclude that we have reached a state on the model side that has no live roles, which is in contradiction with the definition of the model as given in \Cref{fig:evenodd_model}.
Now, for establishing the other part of $\alwaysholds(\xi_{\mathsf{even\_odd}}^{\loc},c,\mstate)$, we follow an approach similar to the two preceding examples.
We use resource predicates $\authfull_{\evenname}(n)$, $\authfrag_{\evenname}(n)$, $\authfull_{\oddname}(n)$, and $\authfrag_{\oddname}(n)$, and we will use the full parts in the invariant below while the fragment parts are passed to the two threads.
The invariant below immediately allows us to establish that $\xi_{\mathsf{match}}^{\loc}$ is preserved throughout program's execution.
\begin{align*}
  & \mathsf{EO\_inv}(\evenname,\oddname,\loc) \triangleq\\
  & \hspace{3em} \Exists n. \fairmodelfrag(n) \ast \loc \mapsto n \ast \left(\mathit{iseven}(n) \ast \authfull_{\evenname}(n) \ast \authfull_{\oddname}(n+1) \right) \lor \left(\mathit{isodd}(n) \ast \authfull_{\evenname}(n+1) \ast \authfull_{\oddname}(n)\right)
\end{align*}
This invariant combines the ideas of the invariants of the preceding two examples.
In particular, it lets us modularly prove that the number stored in $\loc$ corresponds to the model state.
Note how in each case, whether $n$ is even or odd, the number tracked by the odd and even threads, respectively, are out of sync by being one ahead of $n$.
This very closely reflects the behavior of the program; the thread that is out of sync is waiting for the in-sync thread to perform its increment.

\newcommand{\evenrole}{\mathsf{even}}
\newcommand{\oddrole}{\mathsf{odd}}

Finally, we need to prove the weakest precondition of the program.
\[
(\knowInv{\namesp_{\mathit{eo}}}{\mathsf{EO\_inv}(\evenname,\oddname,\loc)} \ast
\hasfuel{\zeta}{
  \lbrace \evenrole := \fuellimit;
  \oddrole := \fuellimit\rbrace} \ast
  \authfrag_{\evenname}(0) \ast
  \authfrag_{\oddname}(1) \wand
\twpre{\mathsf{eo\_start}\ \loc}[\top]{\hasfuel{\zeta}{\emptyset}}[\zeta]
\]
We prove the two threads satisfy the following specifications for any $n$:
\begin{align*}
  (\knowInv{\namesp_{\mathit{eo}}}{\mathsf{EO\_inv}(\evenname,\oddname,\loc)} \ast
  \hasfuel{\zeta}{\lbrace \evenrole := 42 \rbrace} \ast
  \authfrag_{\evenname}(n) \wand &
                                   \twpre{\mathsf{incr\_loop}\ \loc\ n}[\top]
                                   {\hasfuel{\zeta}{\emptyset}}[\zeta]\\
  (\knowInv{\namesp_{\mathit{eo}}}{\mathsf{EO\_inv}(\evenname,\oddname,\loc)} \ast
  \hasfuel{\zeta}{\lbrace \oddrole := 42 \rbrace} \ast
  \authfrag_{\oddname}(n) \wand &
                                  \twpre{\mathsf{incr\_loop}\ \loc\ n}[\top]
                                  {\hasfuel{\zeta}{\emptyset}}[\zeta]
\end{align*}
The proof follows similarly to the one of \Cref{sec:example_yesno}.
We similarly open the invariant around the $\langkw{cas}$ operation and in either case, whether it succeeds or fails, we take the appropriate model step.

\subsection{Obtaining the \Fairis Adequacy Theorem}
\label{sec:fairis_adequacy}

This section describes in more detail how the \Fairis logic and its adequacy theorem are obtained.
In particular, we describe the critical $\Fuel$-construction which, given a model~$\model$, yields a \Trillium{} model~$\Fuel(\model)$ that handles fairness preservation and finite stuttering.
The \Fairis{} logic is then obtained by instantiating \Trillium{} with this model and a well-chosen trace interpretation predicate~$\stateinterp$ and relation~$\fairrel$.

\paragraph{The $\Fuel$-construction}

%
Recall that a state~$\lstate$ of~$\Fuel(\fmodel)$ has two components: a state
$\lstate.\ustate$ of the fairness model~$\fmodel$, and a partial
function~$\lstate.\themap: \mathsf{ThreadId} \rightharpoonup
(\mathsf{Role(\genmod)} \rightharpoonup \nat)$ associating roles to thread ids
and fuels (natural numbers) to roles.
In actuality, not all such pairs are states of~$\Fuel(\fmodel)$, as states
must satisfy certain conditions.
Each role must be associated to at most one thread id:
\[
  \forall \zeta_1 \neq \zeta_2.\;\; \dom(\lstate.\themap(\zeta_1)) \cap
  \dom(\lstate.\themap(\zeta_2)) \;=\; \emptyset
\]
and each \emph{live role} must be associated to a thread id:
\[
  \forall \rho \in \liveroles(\lstate.\ustate).\;
  \exists \zeta \in \mathsf{ThreadId}.\;\;
  \rho \in \dom(\lstate.\themap(\zeta))
\]
where the live roles~$\liveroles(\mstate)$ of a state~$\mstate \in
\model$ are defined as the set $\{ \rho \mid \exists \mstate',\; \mstate
\lto{\rho} \mstate' \}$ of roles that can take a step from~$\mstate$.
Intuitively, to prove fairness of a model trace which refines a fair program execution, we need to show that any live role eventually takes a step, so we need to relate \emph{all} live roles to locales.
We write~$\rdom(\lstate)$ for the set of roles allocated in~$\lstate.\themap$,
formally, $\uplus_{\zeta\in\dom(\lstate.\themap)} \dom(\lstate.\themap(\zeta))$.

Roles that are in $\rdom(\lstate) \setminus \liveroles(\lstate.\ustate)$ have a specific purpose: threads must be associated to a role to be allowed to take steps.
It is often the case that a thread has logically terminated (in that it took all the steps corresponding to steps in the model) but has not actually terminated.
This flexibility allows the thread to terminate in a bounded number of steps without cluttering the model.

\begin{figure}[t]
  \small
  \begin{center}
    \begin{tabular}{l|l|l}
      & \hfil $\lstate \lto{\;\;\livetau{\locale}\;\;} \lstate'$
      & \hfil $\lstate \lto{\;\;\livevis{\locale}{\rho_t}\;\;} \lstate'$ \\
      \hline
      \textsc{st} &
      $\lstate.\ustate = \lstate'.\ustate$ &
      $\lstate.\ustate \lto{\;\;\rho\;\;} \lstate'.\ustate$ \\
      \textsc{th} &
      $\locale \in \rng(\threadmap)$ &
      $\threadmap(\rho) = \locale$ \\
      \textsc{dom} &
      $\rdom(\lstate') \subseteq \rdom(\lstate)$ &
      $\rdom(\lstate') \setminus \rdom(\lstate) \subseteq
      \liveroles(\lstate') \setminus \liveroles(\lstate)$
      \\
      \textsc{dec} &
      $\forall \rho \in \dom(\lstate.\themap(\locale)),\; \fuelmap'(\rho) < \fuelmap(\rho)$ &
      $\forall \rho \in \dom(\lstate.\themap(\locale)) \setminus \{ \rho_t \},\; \fuelmap'(\rho) < \fuelmap(\rho)$
      \\
      \textsc{ch} &
      $\forall \rho,\; \threadmap'(\rho) \neq \threadmap(\rho) \;\Rightarrow\;
      \fuelmap'(\rho) < \fuelmap(\rho)$ &
      $\forall \rho \neq \rho_t,\; \threadmap'(\rho) \neq \threadmap(\rho) \;\Rightarrow\;
      \fuelmap'(\rho) < \fuelmap(\rho)$
      \\
      \textsc{ni} &
      $\forall \rho \notin \dom(\lstate.\themap(\locale)),\;\; \fuelmap'(\rho) \leq \fuelmap(\rho)$ &
      $\forall \rho \notin \dom(\lstate.\themap(\locale)),\;\; \fuelmap'(\rho) \leq \fuelmap(\rho)$
      \\
      \textsc{ref} &&
      $\forall \rho \in \rdom(\lstate') \setminus \rdom(\lstate) \cup \{\rho_t\},\;
      \fuelmap'(\rho) \leq \fuellimit$
    \end{tabular}
  \end{center}
  \caption{Transitions between two states $\lstate$ and~$\lstate'$
    in~$\Fuel(\fmodel)$ where
    $\threadmap \eqdef \threadmap_{\lstate}, \threadmap' \eqdef \threadmap_{\lstate'}$, etc.
    All conditions in the column must be satisfied for the transition in the top row to be valid.
  }
  \label{fig:live-transitions}
\end{figure}

We now turn to the more delicate aspect of the $\Fuel$-construction: the
transitions.
There are two types of transitions: silent stuttering steps are labeled with
$\livetau{\locale}$, where $\locale$ is meant to be the locale of the thread
that takes the stuttering step, and visible steps, corresponding to the
underlying model steps which are labeled with $\livevis{\locale}{\rho}$
annotated with the local $\zeta$ taking the step, and the underlying role $\rho$
in~$\fmodel$.
The locales in labels are necessary to relate $\Fuel(\fmodel)$ steps with the map~$\lstate.\threadmap$.

\Cref{fig:live-transitions} presents the list of conditions that define each
type of transition. The conditions are stated using two auxiliary maps
$\fuelmap_\lstate: \mathsf{Roles} \rightharpoonup \NN$ associating role with its
fuel, and $\threadmap_\lstate: \mathsf{Roles} \rightharpoonup \mathsf{ThreadId}$
associating each role to some thread id.
These two maps are derived from $\lstate.\themap$ as follows:
\[
  \fuelmap_\lstate(\rho) = \fuel \iff \exists \zeta, \fuelsvar.\; \lstate.\themap(\zeta) =
  \fuelsvar \land \fuelsvar(\rho) = \fuel
  \qquad
  \threadmap_\lstate(\rho) = \zeta \iff \rho \in \dom(\lstate.\themap(\zeta))
\]

Condition \textsc{st} restricts the evolution of the underlying fairness model,
\textsc{th}~relates the thread with the state; \textsc{dom} restricts the
evolution of the domain: the only new roles are the roles which became live in
the new underlying state (if any); \textsc{dec} states that all threads
associated to the thread that took the step must decrease their fuel, except
possibly for the role that took a step, \textsc{ni} states that roles that
changed theads must decrease their fuel; \textsc{ni} states that all other roles
cannot increase their fuel, and finally \textsc{ref} restricts the new fuel of
the new roles and of the role that took the step to the global bound~$\fuellimit$.
This last condition is necessary to restrict the branching of the model
$\Fuel(\fmodel)$, to be able to use the adequacy theorem of \Trillium.
It may not be obvious why condition~\textsc{ch} is necessary: otherwise two
threads could, at each step, take visible steps with roles~$\rho_1$ and~$\rho_2$
and, during that step, and change ownership of~$\rho_3$, whose fuel could remain
constant according to condition~\textsc{nd}.

\paragraph{Logical Resources and the Trace Interpretation}

\newcommand{\FM}{\mathit{FM}}
\newcommand{\FuelMap}{\mathit{FuelMap}}
\newcommand{\threadroles}{\mathit{ThreadRoles}}
\newcommand{\rolefuel}{\mathit{RoleFuel}}

The logical predicate $\hasfuel{\locale}{\fuelsvar}$ reflects a coherent
view of the data in the $\Fuel(\fmodel)$ model, as explained below.

To define the $\hasfuel{}{}$ predicate we use Iris's support for custom
ghost state.
These predicates satisfy, among other things, the following rules:
\begin{mathpar}
  \inferH{FuelMap-agree}{\FuelMap(\FM) \and \hasfuel{\locale}{fs}}{\FM(\locale) = fs}
  \and
  \inferH{FuelMap-update}
  {\FuelMap(\FM) \and \hasfuel{\locale}{fs}}
  {\pvs \FuelMap(\FM[\locale := fs']) \ast \hasfuel{\locale}{fs'}}
\end{mathpar}
The link between the state of the $\Fuel$-model and these resources, as well as between locales recorded in the $\Fuel$-model and the program's locales, is specified in the trace interpretation $\stateinterp(\extr, \modtr)$.
First, $\extr$ and $\modtr$ must have the same size, and their corresponding labels
must match: the $i$th label~$\locale_i$ of $\extr$ is equal
to~$\mathit{locale}(l_i)$. Here~$l_i$ is the $i$th label of~$\modtr$, and
$\mathit{locale}$ extracts the locale from a label of~$\Fuel(\fmodel)$, \ie{},
$\mathit{locale}(\livetau{\zeta}) = \mathit{locale}(\livevis{\zeta}{\rho}) =
\zeta$.
Second, the respective last states $(\threadpool, \progstate)$ and $(\ustate,
\fuelmap)$ of~$\extr$ and~$\modtr$ satisfy the following \Iris predicate:
\begin{align*}
  \Exists \FM. &
  \FuelMap(\FM) \ast \fairmodelfull\,\ustate \ast \mathit{coherent}(\FM, \ustate, \fuelmap) \ast
  \dom(\FM) \subseteq \dom(\threadpool) \ast \mathit{ownHeap}(\progstate)
\end{align*}
where $\mathit{ownHeap}(\progstate)$ is the usual state interpretation of the \Iris{} program logic reflecting the program's heap into \Iris{} resources.
Furthermore, $\mathit{coherent}(\FM, \ustate, \fuelmap)$ is a predicate capturing that the ghost fuel map is coherent with respect to the model
state.
In particular, coherence captures that the
ghost fuel map and model fuel map have the same locales, and
that any role present in the former is also present in the latter with a larger or equal fuel.
Additionally, coherence captures that any live role of the model exists in the ghost fuel map.

\paragraph{Adequacy}

To use \Trillium's adequacy theorem, the last missing piece is to choose the relation~$\fairrel = \Fuel(\xi)$.
It turns out that, for our purpose, this relation can be rather weak.
Given a finite program execution~$\extr = (\threadpool_1, \progstate_1)
\xrightarrow{\locale_1} (\threadpool_2, \progstate_2) \xrightarrow{\locale_2} \cdots \xrightarrow{\locale_{n-1}} (\threadpool_n,
\progstate_n)$, and a finite $\Fuel(\fmodel)$-trace $\modtr = (\ustate_1,
\themap_1) \xrightarrow{l_1} (\ustate_2, \themap_2) \xrightarrow{l_2} \cdots \xrightarrow{l_{n-1}} (\ustate_n,
\themap_n)$, we define $\fairrel(\extr, \modtr)$ as follows:
\begin{align}
  \fairrel(\extr, \modtr) & \eqdef{} \xi(\extr, \modtr) \land{} \notag \\
  & \left(\forall j,\; \locale_j =
  \mathit{locale}(l_j)\right) \label{xi-fair-two} \land{}\\
  &\left(\forall i,\; \rng(\threadmap_{\lstate_i}) \subseteq \dom(\threadpool_i) \land
    \forall \locale, \role,\; \threadmap_{\lstate_i}(\role) = \locale
  \text{$\threadpool_i[\locale]$ is a value} \Rightarrow \rho \notin
\liveroles(\ustate_i) \right) \label{xi-fair-one}
\end{align}
We can now use \Trillium's adequacy theorem (\cref{thm:adequacy}) to prove Fairis's adequacy theorem (\cref{thm:fairis_adequacy}) given the following lemma.
\begin{lemma} The following holds for the parameters we have chosen:
  \[ \alwaysholds(\Fuel(\xi), c, \initialfuelstate{m}) \implies \alwaysholds(\xi, c, m) \]
\end{lemma}
\begin{proof}[Proof sketch]
  Let us consider a program execution~$\extr$ and a model trace~$\modtr$.
  It suffices to consider their last states, which we write
  respectively as $(\threadpool, \progstate)$ and $\lstate$.
  That $\rng(\threadmap_\lstate) \subseteq \dom(\threadpool)$ follows from
  $\mathit{coherent}(\FM, \lstate.\ustate, \fuelmap_\lstate)$ and from the fact that $\dom(\threadmap_\lstate)
  = \dom(\fuelmap_\lstate)$.
  We prove the contrapositive of the second conjunct of \eqref{xi-fair-one} above.
  Assume a locale $\locale$ and a role $\role$ such that $\threadmap_\lstate(\role) =
  \locale$ and such that~$\threadpool[\locale]$ is a value. Since we know all
  the postconditions of terminated threads hold, we know
  $\hasfuel{\locale}{\emptyset}$, which means, according to
  $\mathit{coherent}(\FM, \lstate.\ustate, \fuelmap_\lstate)$, that $\role$ cannot be live in
  $\lstate.\ustate$.
  The condition \eqref{xi-fair-two} on labels follows directly from the trace interpretation.
\end{proof}

\paragraph{Fairness Preservation}

The adequacy theorem of \Fairis gives us that the intial states of the program
and of the $\Fuel(\model)$ model are related: $\singletontrace{(\expr, \progstate)}
\continuedsim{\fuelxi} \singletontrace{\initialfuelstate{\mstate}}$.
This refinement is useful because it has good properties: preservation of
fairness and preservation of termination.
To make things precise, we define some operations on traces:
Any finite or infinite trace~$\modtr$ of~$\Fuel(\fmodel)$ induces a trace $\destutter{\modtr}$
in~$\fmodel$ which we obtain by removing $\livetau{}$ transitions and projecting
out the $\ustate$ component.
Since there can only be finitely many~$\livetau{}$ transitions in a row (any
such transition decreases the sum of all fuels), $\modtr$~is finite if and only
if $\destutter{\modtr}$~is finite. And since~$\hat{\xi}_{\mathit{fuel}}$ relates
traces of the same length, if $\destutter{\modtr}$ is finite and
if~$\fairtrhold{\extr}{\modtr}$ holds, then~$\extr$ is finite.

Let us now explain why fairness is preserved in the other direction:
if~$\fairtrhold{\extr}{\modtr}$ and~$\extr$ is fair, then $\destutter{\modtr}$
is fair. We endow traces of~$\Fuel(\fmodel)$ with the natural notion of fairness
where we only look at the \emph{roles} of the transitions, not the locales. It
is obvious that if $\modtr$ is fair then $\destutter{\modtr}$ is fair, as
$\livetau{}$-transitions do not play any role in fairness.
Therefore, preservation of fairness boils down to:
\begin{lemma}
  If $\extr$ is fair and if~$\fairtrhold{\extr}{\modtr}$, then~$\modtr$ is fair.
\end{lemma}
\begin{proof}[Proof sketch]

The proof is quite technical and consists of two nested inductions, but the idea
is the following: consider some state~$\lstate$ of~$\modtr$ and~$\rho$ which is
live in that state, and call~$\locale = \threadmap_\lstate(\rho)$ its associated locale.
We need to prove there eventually exists a $\rho$ transition in~$\modtr$.
Since we have a fixed fair program trace~$\extr$, there exists~$n$ such that the
next step of~$\locale$ is in at most $n$~steps.
We proceed by induction over $(\fuelmap_\lstate(\rho), n)$, ordered
lexicographically.

Without loss of generality, we can assume it is the first state of~$\modtr$.
Write $\locale$ for~$\threadmap_\lstate(\role)$.
Consider the first step of the program with locale~$\locale'$. There are two
cases: (1) $\locale \neq \locale'$. Then we apply the induction hypothesis
with the next state, the same fuel, and $n-1$.
(2) $\locale = \locale'$. If it is a $\rho$-transition we conclude.
Otherwise, the fuel associated to~$\rho$ decreases, and we apply the
induction hypothesis with the next state, a smaller fuel and some
arbitrary~$n'$ obtained as above.
\end{proof}

\section{The Aneris Logic}\label{sec:aneris}
To reason about distributed systems, we instantiate \Trillium{} with \AnerisLang{}, the programming language accompanying \Aneris{}, a higher-order distributed separation logic \cite{DBLP:conf/esop/Krogh-Jespersen20}.
\AnerisLang{} is an OCaml-like programming language with network primitives for creating (\newsocket) and binding (\socketbind) network sockets as well as sending (\sendto) and receiving (\receivefrom) messages.
The operational semantics of \AnerisLang{} is designed so that the primitives closely model Unix sockets and UDP (unreliable) networking.

The \Aneris{} instantiation of \Trillium{} is conceptually simple as we will target safety trace properties.
This means that we can ``bake-in'' the reflexive closure of the model and hence freely allow model stuttering.
The result is a program logic and reasoning principles that are virtually identical to the original (non-relational) \Aneris{} program logic, where the only difference is the addition of a single rule (\ruleref{aneris-take-step}) that allows us to relate an atomic step of the program to a corresponding step in the LTS model.
\begin{mathpar}
  \inferH
  {aneris-take-step}
  { \twpre{\expr}[\mask]{\val.\; \modelfrag(\genmodstate') \wand \pred(\val)}[\zeta] \\
    \modelfrag(\genmodstate) \\
    \genmodstate \genmodstepof{\genmod} \genmodstate' \and \text{Atomic}(\expr) \and \expr \not\in \Val}
  {\twpre{\expr}[\mask]{\pred}[\zeta]}
\end{mathpar}
While fairly simple, this instantiation will still allow us to prove interesting properties.
In what follows, we will show how we use \Aneris{} to transport safety (trace) properties of \TLA{} protocol models to distributed programs that implement them.
In the Appendix we also show how we, under reasonable liveness assumptions, can prove \emph{eventual consistency} \citep{DBLP:journals/cacm/Vogels09} of a Conflict-Free Replicated Data Type~\cite{DBLP:journals/eatcs/ShapiroPBZ11}.
We leave a more principled approach to proving liveness properties of distributed systems as future work.

We use the \Aneris{} instantiation of \Trillium{} to show an intensional refinement between implementations of two classical distributed algorithms, Two-Phase Commit (TPC) \cite{gray1978notes} and Single-Decree Paxos (SDP) \cite{DBLP:journals/tocs/Lamport98, lamport2001paxos}, and their \TLA{} \cite{lamport1993hybrid} models.
As simple corollaries of the refinement, we show using a \emph{single} modular specification
\begin{enumerate*}
\item that clients are \emph{safe}, \ie{}, they do not crash,
\item a formal proof that the implementation \emph{correctly} implements a protocol, and
\item correctness of the implementation by leveraging existing correctness properties of the models.
\end{enumerate*}
\href{https://github.com/tlaplus/Examples/blob/master/specifications/transaction_commit}{The \TLA{} specification of TPC} and \href{https://github.com/tlaplus/Examples/tree/master/specifications/Paxos}{the \TLA{} specification of SDP} can both be found in the official \TLA-examples repository on GitHub.
In our formalization, we have manually translated the \TLA{} protocol specifications into STSs in \Coq{} and proved their correctness properties.\footnote{A user who does not aim to be as foundational could, however, trust the translation and the existing \TLA{} proofs.}

Note that correctness of the implementations \emph{can} be established using regular \Iris{} ghost resources and invariants, but doing so through a refinement has the immediate benefit that the necessary ghost theory is much simpler.
The protocol logic is already encoded in the model and we ``just'' need to map the state of the model to the physical state of the distributed system.
The only place we will need more sophisticated ghost theory is where the model is underspecified, \eg{}, in how SDP distributes ballots among proposers.
Additionally, by showing an intensional refinement, we show that the implementation of, say, SDP \emph{actually} implements the SDP protocol and not just any other consensus protocol.
While we do not show this explicitly, it also means that it is possible to transfer other trace properties of the model to the implementation.

Both the implementation, the model, and the refinement proof for the TPC
protocol can be found in the Appendix.
The development follows the same methodology as for SDP, which we describe below; we omit network- and state-related \Aneris{} resources and focus on the core parts relevant for showing the refinement.

\paragraph{Single-Decree Paxos}

The Paxos algorithm is a consensus protocol and its single-decree version allows a set of distributed nodes to reach agreement on a single value by communicating through message-passing over an unreliable network.

In SDP, each node in the system adopts one or more of the responsibilities of either \emph{proposer}, \emph{acceptor}, or \emph{learner}.
A value is chosen when a learner learns that a \emph{quorum} (\eg, a majority) of acceptors have accepted a value proposed by some proposer.
The algorithm works in two phases: in the first phase, a proposer tries to convince a quorum of acceptors to promise that they will later accept its value.
If it succeeds, it continues to the second phase where it asks the acceptors to fulfill their promise and accept its value.
To satisfy the requirements of consensus, each attempt to decide a value is distinguished with a unique totally-ordered round number or \emph{ballot}.
Each acceptor stores its current ballot and the last value it might have accepted, if any.
Acceptors will only give a promise to proposers with a ballot greater than their current one, and in that case they switch to the proposer's ballot; proposers only propose values that ensure consistency, if chosen.
By observing that a quorum of acceptors have accepted a value for the same ballot, learners will learn that a value has been chosen.
We refer to \citet{lamport2001paxos} for an elaborate textual description of the protocol.

\paragraph{Model}
The \TLA{} model of SDP is summarized in \cref{fig:paxos_model}.
The model is parameterized over a set of acceptors, $\paxosAcceptors$, and a type of values, $\paxosValue$, among which values are chosen.
The state of the model consists of a set of sent messages $\paxosmsgs \in \pset{\paxosMdlMessages}$ and two maps $\paxosMaxBal : \paxosAcceptors \ra \option{\paxosBallot}$ and $\paxosMaxVal : \paxosAcceptors \ra \option{\paxosBallot \times \paxosValue}$ that for each acceptor record the greatest ballot promise and the last accepted value together with its ballot, respectively.
The message type is defined using a datatype-like notation as
\[
  \paxosMdlMessages \eqdef{}
                      \paxosmsgOneA{\paxosbal} \mid
                      \paxosmsgOneB{\paxosacceptor}{\paxosbal}{\paxosmval} \mid
                      \paxosmsgTwoA{\paxosbal}{\paxosval} \mid
                      \paxosmsgTwoB{\paxosacceptor}{\paxosbal}{\paxosval}
\]
where $\paxosacceptor \in \paxosAcceptors$, $\paxosbal \in \paxosBallot$, $\paxosval \in \paxosValue$, and $\paxosmval \in \option{\paxosBallot \times \paxosValue}$.

\begin{figure}[htb!]
  \centering
  \begin{align*}
    Q1bv(\paxosmsgs, Q, \paxosbal)
    \eqdef{}& \set{ m \in \paxosmsgs \mid \Exists \paxosacceptor, \paxosval . m = \paxosmsgOneB{\paxosacceptor}{\paxosbal}{\Some{\paxosval}} \land \paxosacceptor \in Q } \\
    \paxosHavePromised(\paxosmsgs, Q, \paxosbal)
    \eqdef{}& \All a \in Q . \Exists m \in \paxosmsgs, \paxosmval . m = \paxosmsgOneB{\paxosacceptor}{\paxosbal}{\paxosmval}   \\
    \paxosIsMaxVote(\paxosmsgs, Q, \paxosbal, \paxosval)
    \eqdef{}& \Exists m_{0} \in Q1bv(\paxosmsgs, Q, \paxosbal), \paxosacceptor_{0}, \paxosbal_{0} .
              m = \paxosmsgOneB{\paxosacceptor_{0}}{\paxosbal}{\Some{\paxosbal_{0}, \paxosval}} \land \\
            &\quad \All m' \in Q1bv(\paxosmsgs, Q, \paxosbal) . \\
            &\qquad \Exists \paxosacceptor', \paxosbal', \paxosval' .
              m' = \paxosmsgOneB{\paxosacceptor'}{\paxosbal}{\Some{\paxosbal', \paxosval'}} \land \paxosbal_{0} \geq \paxosbal' \\
    \paxosShowsSafeAt(\paxosmsgs, Q, \paxosbal, \paxosval)
    \eqdef{}& \paxosHavePromised(\paxosmsgs, Q, \paxosbal) \; \land
            \left( Q1bv(\paxosmsgs, Q, \paxosbal) = \emptyset\spac \lor \paxosIsMaxVote(\paxosmsgs, Q, \paxosbal, \paxosval) \right)
  \end{align*}
  \begin{mathparpagebreakable}
    \small
    \inferH{SDP-Phase1a}
    {\phantom{A}}
    {\paxosstep
      {\paxosmsgs, \paxosMaxBal, \paxosMaxVal}
      {\paxosmsgs \cup \set{\paxosmsgOneA{\paxosbal}}, \paxosMaxBal, \paxosMaxVal}
    }
    \and
    \inferH{SDP-Phase1b}
    { \paxosmsgOneA{\paxosbal} \in \paxosmsgs \\
      \paxosbal > \paxosMaxBal(\paxosacceptor) \\
      \paxosMaxVal(\paxosacceptor) = \paxosmval
    }
    {\paxosstep
      {\paxosmsgs, \paxosMaxBal, \paxosMaxVal}
      {\paxosmsgs \cup \set{\paxosmsgOneB{\paxosacceptor}{\paxosbal}{\paxosmval}},
        \paxosMaxBal[\paxosacceptor \mapsto \Some{\paxosbal}],
        \paxosMaxVal}
    }
    \and
    \inferH{SDP-Phase2a}
    { \not\exists \paxosval' .\spac \paxosmsgTwoA{\paxosbal}{\paxosval'} \in \paxosmsgs \\
      \paxosQuorum(Q) \\
      \paxosShowsSafeAt(\paxosmsgs, Q, \paxosbal, \paxosval)
    }
    {\paxosstep
      {\paxosmsgs, \paxosMaxBal, \paxosMaxVal}
      {\paxosmsgs \cup \set{\paxosmsgTwoA{\paxosbal}{\paxosval}}, \paxosMaxBal, \paxosMaxVal}
    }
    \and
    \inferH{SDP-Phase2b}
    { \paxosmsgTwoA{\paxosbal}{\paxosval} \in \paxosmsgs \\
      \paxosbal \geq \paxosMaxBal(\paxosacceptor)
    }
    {\paxosstep
      {\paxosmsgs, \paxosMaxBal, \paxosMaxVal}
      {\paxosmsgs \cup \set{\paxosmsgTwoB{\paxosacceptor}{\paxosbal}{\paxosval}},
        \paxosMaxBal[\paxosacceptor \mapsto \Some{\paxosbal}],
        \paxosMaxVal[\paxosacceptor \mapsto \Some{\paxosbal, \paxosval}]}
    }
\end{mathparpagebreakable}
\caption{\TLA\ specification of single-decree Paxos ($\mathsf{SDP}$).}
\label{fig:paxos_model}
\end{figure}

The \ruleref{SDP-Phase1a} transition adds a $\paxosmsgOneA{\paxosbal}$ message to the set of sent messages; this corresponds to the proposer asking the acceptors to \emph{not} accept values for ballots smaller than $\paxosbal$.
If a $\paxosmsgOneA{\paxosbal}$ message has been sent and $b$ is greater than acceptor $\paxosacceptor$'s current ballot $\paxosMaxBal(\paxosacceptor)$ then the \ruleref{SDP-Phase1b} transition updates $\paxosacceptor$'s state and sends a $\paxosmsgOneB{\paxosacceptor}{\paxosbal}{\paxosmval}$ message where $\paxosmval$ is $\paxosacceptor$'s last accepted value, if any.
This corresponds to an acceptor responding to a proposer's promise request.

The second phase is initiated using the \ruleref{SDP-Phase2a} transition that corresponds to the proposer proposing a value $\paxosval$ for ballot $\paxosbal$ by sending a $\paxosmsgTwoA{\paxosbal}{\paxosval}$ message.
However, the transition can only be made if no value has previously been proposed for ballot $\paxosbal$ and if a quorum $Q$ of acceptors exists such that the $\paxosShowsSafeAt(\paxosmsgs, Q, \paxosbal, \paxosval)$ predicate holds; this predicate is at the heart of the Paxos algorithm.
Intuitively, the predicate holds if all acceptors in $Q$ have promised not to accept values for any ballot less than $\paxosbal$ ($\paxosHavePromised(\paxosmsgs, Q, \paxosbal)$) and \emph{either} none of the acceptors have accepted any value for all ballots less than $\paxosbal$ \emph{or} $\paxosval$ is the value of the largest ballot that acceptors from $Q$ have accepted.
Following the \ruleref{SDP-Phase2b} transition, acceptor $\paxosacceptor$ may accept a proposal for value $\paxosval$ and ballot $\paxosbal$ by sending a $\paxosmsgTwoB{\paxosacceptor}{\paxosbal}{\paxosval}$ message and updating its state to reflect this fact.
A value $\paxosval$ has been chosen when a quorum of acceptors have sent a $\paxosmsgTwoB{\paxosacceptor}{\paxosbal}{\paxosval}$ message for some ballot $\paxosbal$:
\[
    \paxosChosen(\paxosmsgs, \paxosval)
    \eqdef{} \Exists \paxosbal, Q . \paxosQuorum(Q) \land \All \paxosacceptor \in Q . \paxosmsgTwoB{\paxosacceptor}{\paxosbal}{\paxosval} \in \paxosmsgs
  \]
  As follows from the theorem below, it is not possible for the protocol to choose two different values at the same time and hence SDP solves the consensus problem.
\begin{theorem}[Consistency, $\SDP$ model]\label{thm:paxos-consistency}
  Let $\paxosinit = (\emptyset, \Lam \_ . \None, \Lam \_ . \None)$.
  If $\paxossteprtc{\paxosinit}{(\paxosmsgs, \paxosMaxBal, \paxosMaxVal)}$ and both $\paxosChosen(\paxosmsgs, \paxosval_{1})$ and $\paxosChosen(\paxosmsgs, \paxosval_{2})$ hold then $\paxosval_{1} = \paxosval_{2}$.
\end{theorem}

\paragraph{Implementation}
\cref{lst:paxos_acceptor} and \cref{lst:paxos_proposer} show implementations of the acceptor and proposer roles, respectively.
The learner implementation and utility functions such as
$\textlang{recv\_promises}$ and $\textlang{find\_max\_promise}$ are found in the
Appendix.

\begin{figure}[t!]
  \centering
  \begin{minipage}[t]{0.54\linewidth}
\begin{AnerisPL}[label={lst:paxos_acceptor}, caption={Acceptor implementation.}]
let acceptor learners addr =
  let skt = socket () in
  socketbind skt addr;
  let maxBal = ref None in
  let maxVal = ref None in
  let rec loop () =
    let (m, sndr) = receivefrom skt in
    match acceptor_deser m with
    | inl bal =>
      if !maxBal = None ||
         Option.get !maxBal < bal then
        maxBal := Some bal;
        sendto skt
          (proposer_ser (bal, !maxVal)) sndr
      else ()
    | inr (bal, v) =>
      if !maxBal = None ||
         Option.get !maxBal <= bal then
        maxBal := Some bal;
        maxVal := Some accept;
        sendto_all skt learners
          (learner_ser (bal, v))
      else ()
    end; loop () in loop ()
\end{AnerisPL}
  \end{minipage}
  \begin{minipage}[t]{0.45\linewidth}
\begin{AnerisPL}[label={lst:paxos_proposer}, caption={Proposer implementation.}]
let proposer acceptors skt bal v =
  sendto_all skt acceptors
    (acceptor_ser (inl bal));
  let majority =
    (Set.cardinal acceptors) / 2 + 1 in
  let promises =
    recv_promises skt majority bal in
  let max_promise =
    find_max_promise promises in
  let av = Option.value max_promise v in
  sendto_all skt acceptors
    (acceptor_ser (inr (bal, av)))
\end{AnerisPL}
\begin{AnerisPL}[label={lst:paxos_client}, caption={Client implementation.}]
let client addr =
  let skt = socket () in
  socketbind skt addr;
  let (m1, sndr1) = receivefrom skt in
  let (_, v1) = client_deser m1 in
  let (m2, _) = wait_receivefrom skt
    (fun (_, sndr2), sndr2 <> sndr1) in
  let (_, v2) = client_deser m2 in
  assert (v1 = v2); v1.
\end{AnerisPL}
  \end{minipage}
\end{figure}

The acceptor implementation receives as input a set of learner socket addresses and an address to communicate on.
It creates a fresh socket, binds it to the address, and allocates two local references to keep track of its current ballot and last accepted value.
In a loop, it listens for the two different kinds of messages that it may receive from the proposers.
Given a phase one message, it only considers the message if the ballot is greater than its current ballot in which case it responds with its last accepted value.
Given a phase two message, it only considers the message if the ballot is greater than or equal to its current ballot in which case it accepts the value and broadcasts the fact to all the learners.
The learner implementation (included in the Appendix) simply waits for such a message for the same ballot from a majority of acceptors.

The proposer implementation receives as input a set of acceptor socket addresses, a bound socket, a ballot number and a value to (possibly) propose in the ballot.
First phase is initiated by sending a message to all the acceptors and after receiving a response from a majority of the acceptors it continues to the second phase.
In the second phase it picks the value of the maximum ballot among the responses; if no such value exist, it picks its own.
The candidate is finally sent to all acceptors.

Note that this proposer implementation only proposes a value for a single ballot; typically, proposers will issue new ballots when learning that no decision has been reached due to messages being dropped or nodes crashing.
Moreover, it is crucial that proposers do not issue proposals for the same ballot.
In our Coq formalization, proposer $\paxosproposer$ repeatedly issues new ballots of the form $k \cdot |\paxosProposers| + \paxosproposer$ for $k \in \mathbb{N}$ by keeping track of the last issued $k$ in a local reference.

\paragraph{Consensus by Refinement}

To show that the SDP implementation refines the SDP model we instantiate the \Aneris~logic with the model; the key part of the proof is to keep the $\modelfrag(\paxosstate)$ resource in a global invariant that ties together the model state and the physical state with enough information to verify the implementation and for the refinement relation established through the adequacy theorem to be strong enough for proving our final correctness theorem (\cref{cor:paxos-impl-consensus}).
Under this invariant we will \emph{modularly} verify each Paxos role and each component in isolation.

\sloppy To state the invariant, we use three kinds of resources corresponding to:
\begin{enumerate}
\item sets of messages with predicates $\paxosmsgauth(\paxosmsgs)$ and $\paxosmsgfrag(m)$ such that
  \begin{align*}
    \paxosmsgauth(\paxosmsgs) \asts \paxosmsgfrag(m) &\proves m \in \paxosmsgs \\
    \paxosmsgauth(\paxosmsgs) &\proves \pvs \left(\paxosmsgauth(\paxosmsgs \cup m) \asts \paxosmsgfrag(m)\right)
  \end{align*}
\item maps, \eg, with predicates $\paxosmbauth(\paxosMaxBal)$ and $\paxosmbfrag(\paxosacceptor, \paxosbal)$ such that
  \begin{align*}
    \paxosmbauth(\paxosMaxBal) \asts \paxosmbfrag(\paxosacceptor, b) &\proves \paxosMaxBal(\paxosacceptor) = b \\
    \paxosmbauth(\paxosMaxBal) \asts \paxosmbfrag(\paxosacceptor, b) &\proves \pvs \left(\paxosmbauth(\paxosMaxBal[\paxosacceptor \mapsto b']) \asts \paxosmbfrag(\paxosacceptor, b')\right)
  \end{align*}
\item ballots with predicates $\ospending(\paxosbal)$ and $\osshot(\paxosbal, \paxosval)$ such that \\
  \begin{minipage}{0.4\linewidth}
    \begin{align*}
      \ospending(\paxosbal) \asts \osshot(\paxosbal, \paxosval) &\proves \FALSE \\
      \ospending(\paxosbal) \asts \ospending(\paxosbal) &\proves \FALSE
    \end{align*}
  \end{minipage}
  \begin{minipage}{0.5\linewidth}
    \begin{align*}
      \ospending(\paxosbal) &\proves \pvs \osshot(\paxosbal, \paxosval) \\
      \osshot(\paxosbal, \paxosval_{1}) \asts \osshot(\paxosbal, \paxosval_{2}) &\proves \paxosval_{1} = \paxosval_{2}
    \end{align*}
  \end{minipage}
  \vspace{0.6em}
\end{enumerate}
Equipped with these resource we can state the invariant:
\begin{align*}
  I_{\SDP}
    &\eqdef{}
      \Exists \paxosmsgs, \paxosMaxBal, \paxosMaxVal .
    \begin{aligned}[t]
      &\modelfrag(\paxosmsgs, \paxosMaxBal, \paxosMaxVal) \asts
      \paxosmsgauth(\paxosmsgs) \asts
      \paxosmbauth(\paxosMaxBal) \asts \\
      & \paxosmvauth(\paxosMaxVal) \asts
      \paxosbalcoh(\paxosmsgs) \asts
      \paxosmsgcoh(\paxosmsgs)
    \end{aligned}
\end{align*}

The first part of the invariant ties the current state of the model $(\paxosmsgs, \paxosMaxBal, \paxosMaxVal)$ to its logical \emph{authoritative} counterparts which means that by owning a \emph{fragmental} part you own a piece of the model: \eg{}, by owning $\paxosmbfrag(\paxosacceptor, \paxosbal)$ you may open the invariant and conclude $\paxosMaxBal(\paxosacceptor) = \paxosbal$ where $\paxosMaxBal$ is the current map of ballots.
Intuitively, we will give acceptor $\paxosacceptor$ exclusive ownership of the parts of the model that should correspond to its local state (through resources $\paxosmbfrag(\paxosacceptor, \paxosbal)$ and $\paxosmvfrag(\paxosacceptor, \paxosmval)$). 
Similarly, by owning $\paxosmsgfrag(m)$ one may conclude that the message $m$ has in fact been added to the set of messages in the model; this predicate we will transfer when sending physical messages corresponding to $m$.

In the last part of the invariant, the $\paxosbalcoh(\paxosmsgs)$ predicate simply requires that if $\paxosmsgTwoA{\paxosbal}{\paxosval} \in \paxosmsgs$ then $\osshot(\paxosbal, \paxosval)$ holds.
This implies that by owning $\ospending(\paxosbal)$ you are the only entity that may propose a value for ballot $\paxosbal$ and it may never change.
The $\paxosmsgcoh(\paxosmsgs)$ predicate ties the physical state of the program to the model using \Aneris{}-specific predicates for tracking the state of the network.
This, for instance, forces acceptors and proposers to also add to the model state $\paxosmsgs$ any message they send over the network.
Hence, to verify a proposer or an acceptor that sends a message, the proof must open the invariant, use \ruleref{aneris-take-step} to take a step in the model, and update the corresponding logical resources to close the invariant.
Following this methodology, we give specifications of the following shape to the proposer and acceptor components:
\begin{align*}
  &\knowInv{}{I_{\SDP}} \asts \paxosmbfrag(\paxosacceptor, \None) \asts \paxosmvfrag(\paxosacceptor, \None) \asts \ldots \wand{} \twpre{\textlang{acceptor}~L~\paxosacceptor}[\top]{\FALSE}[\zeta] \\
  &\knowInv{}{I_{\SDP}} \asts \ospending(\paxosbal) \asts \ldots \wand \twpre{\textlang{proposer}~A~skt~\paxosbal~\paxosval}[\top]{\TRUE}[\zeta]
\end{align*}
omitting Aneris-specific network connectives in the precondition; the postcondition for $\textlang{acceptor}$ may be $\FALSE$ as it does not terminate.
We give a similar specification to the learner.
Working in a modular program logic, we can compose these specifications to get a single specification for a distributed system with both proposers, acceptors, and learners.
By applying the adequacy theorem to this specification we get that the implementation indeed refines the \TLA{} model of SDP.\footnote{The full Coq proof amounts to about 1100 lines of proof scripts.}

\paragraph{Consensus for the Implementation}
Given the specification has been established for the implementation, we can state and prove that the consistency property holds for all executions by transporting the consistency property of the model. Let
\[
  \ChosenI(M, \paxosval) \eqdef{} \Exists \paxosbal, Q .
  \paxosQuorum(Q) \land
  \All \paxosacceptor \in Q .
  \Exists m \in M . m \sim \paxosmsgTwoB{\paxosacceptor}{\paxosbal}{\paxosval}
\]
where $\ms$ is a set of physical messages and $m \sim s$ holds when $m$ is the serialization of the model message $s$.
By picking a trace relation $\progmodrel_{\SDP}$ that requires messages in the model to correspond to messages in the program state (as implied by $\paxosmsgcoh(\paxosmsgs)$):
\[
  \progmodrel_{\SDP}(\finprogtrace, \finmodtrace) \eqdef{}
  \Exists \paxosmsgs . \tracelast(\finmodtrace) =  (\paxosmsgs, \_, \_) \land
  \MessagesOf(\tracelast(\finprogtrace)) \sim \paxosmsgs \land \stuttering(\finmodtrace)
\]
we combine the adequacy theorem (\cref{thm:adequacy}) with our 
model correctness theorem (\cref{thm:paxos-consistency}) to obtain the following corollary that \emph{only} talks about the execution of the SDP implementation.
\begin{corollary}\label{cor:paxos-impl-consensus}
  Let $\expr$ be a distributed system obtained by composing $n$ proposers, $m$ acceptors, and $k$ learners.
  For any $\tpool$ and $\pstate$, if $(\expr; \emptyset) \step^{*} (\tpool; \pstate)$ and both $\ChosenI(\MessagesOf(\pstate), \val_{1})$ and $\ChosenI(\MessagesOf(\pstate), \val_{2})$ hold then $\val_{1} = \val_{2}$.
\end{corollary}

\paragraph{Functional Correctness}\label{sec:paxos-clients}

\cref{cor:paxos-impl-consensus} is a meta-logic theorem (\eg, in \Coq) that only talks about the program execution and it follows from the adequacy theorem and the model correctness theorem.
However, it is not only in the meta-logic that we can exploit properties of the model to prove properties about programs as the model is also embedded as a \emph{resource} in the logic.

\cref{lst:paxos_client} shows a client application that receives a message from two different Paxos learners and asserts that the two values are equal; if the two values do not agree, the program crashes.
We can prove a specification for the client of the shape $\knowInv{}{I_{\SDP}} \asts \ldots \wand \twpre{\textlang{client}~a}{\ldots}$.
From the adequacy theorem it follows that the program is safe, \ie, it does not crash, which means the asserted statement must always hold.
In the proof of this specification, the client will receive ghost resources from the learners conveying that $\val_{1}$ and $\val_{2}$ have been chosen (\ie, that a quorum of acceptors have accepted $\val_{i}$).
By opening the invariant $I_{\SDP}$ and hence obtaining the model resource $\modelfrag(\paxosmsgs, \paxosMaxBal, \paxosMaxVal)$, we can combine this knowledge with \cref{thm:paxos-consistency}---a property exclusively of the model---and conclude that $\val_{1} = \val_{2}$.
Naturally, we may still compose a distributed system containing the client together with proposers, acceptors, and learner nodes and derive a specification for the full system.
This single specification for the full distributed system entails \emph{both} the refinement of the \TLA{} model and the safety of the programs running on all nodes.

\section{Related Work}
\label{sec:related-work}

\newcommand{\IronFleet}{IronFleet}
\newcommand{\Igloo}{Igloo}
\newcommand{\Dafny}{Dafny}

\paragraph{Refinement-based Verification of Distributed Systems.}
We focus on works that, as ours, aim at proving that concrete implementations refine models.
The most closely related works are \IronFleet{} \citep{10.1145/3068608} and \Igloo{} \citep{10.1145/3428220}.
In contrast to both, our approach is foundational: the operational semantics of the languages, the models, and the program logics are all formally defined in \Coq{}, and through adequacy theorems of the program logic, the end result of a verification is a formal theorem expressed only in terms of the operational semantics of the programming language and the model.

\IronFleet{} uses the \Dafny{} verifier to verify implementations and encodes the refinement of an STS in preconditions and postconditions of programs but does not support node-local concurrency.
\IronFleet{} uses a pen-and-paper argument for proving liveness of simple programs (programs that consist of a simple event loop which calls terminating event handlers).

\Igloo{} proves a particular kind of extensional safety properties.
\Igloo{} refines a high-level STS to a more low-level STS for each node of the system.
STSs are annotated with IO operations which are used to generate IO specifications for network communications of the node in the style of \citet{DBLP:conf/esop/Penninckx0P15}.
Programs are subsequently verified against this generated specification.
The relationship between the implementation and the model considered in \Igloo{} is a fixed relation, \ie{} producing the same IO behavior.
In contrast, our work allows an arbitrary intensional refinement relation to be specified and established between the program and the model.

\paragraph{Refinement in \Iris{}.}
There has been earlier work on proving contextual refinements using \Iris{} as discussed in the introduction.
Additionally, Perennial \cite{DBLP:conf/sosp/ChajedTKZ19} defines correctness of a system using \emph{concurrent recovery refinement}, requiring that the (possibly crashing) implementation and specification STS has the same external I/O.
This notion of refinement is much coarser and does not allow you to prove, \eg{}, fair termination.
\citet{DBLP:journals/pacmpl/TassarottiH19} relates concurrent probabilistic programs to abstract specifications denoting indexed valuations, exhibiting a probabilistic coupling when assuming that the implementation terminates.

Our approach to termination-preserving refinement is similar in spirit to the one of \citet{tassarotti-17}
but applies to reasoning about refinement of general concurrent programs with respect to abstract models, not just compilation of session-typed programs.
To the best of our knowledge, the expressiveness of the logics is roughly similar.
The main difference is that \citet{tassarotti-17} augments the Iris base logic with \emph{linear} propositions, which requires modifying the definition of resource algebra to add a transition relation.
We achieve similar results without heavy modifications, using that the authoritative state of the model is threaded through the weakest precondition, and by putting an exclusive structure on the set of roles owned by a thread, which prevents arbitrary weakening of the role resource, a limited form of linearity.

Besides efforts in Iris, \citet{DBLP:journals/pacmpl/LiangF18,DBLP:conf/popl/LiangF16} have also used refinement to show a wider range of liveness properties of concurrent programs, including programs with partial methods, but focusing on first-order logic and first-order programs.
It would be interesting to investigate if \Trillium{} could serve as a basis for generalizing the verification methods of \citet{DBLP:journals/pacmpl/LiangF18,DBLP:conf/popl/LiangF16} to higher-order logic and higher-order programs.

Simuliris \cite{simuliris} is a separation logic for fair termination-preserving contextual refinements for concurrent program transformations that can exploit undefined behavior.
In contrast to both Iris and Trillium, Simuliris is \emph{not} step-indexed, and thus does not support impredicative invariants or higher-order ghost state, which we crucially target and rely on.\footnote{For example, higher-order ghost state is used to define Aneris's socket protocols.}

\paragraph{Certified Abstraction Layers}

A related approach to verification is \emph{certified abstraction layers}
\citep{cal}, in particular their concurrent variant \citep{ccal}, which are used
to verify the CertiKOS verified kernel \citep{certikos}.
Our approach is similar to that of CertiKOS in that both approaches use models to help
verify programs. The main difference is that, in the CertiKOS approach, the
person proving a refinement needs to work directly using the semantics of the
program and of the model, which are both sets of traces. Our approach, on the other
hand, is to use a program logic to do the heavy lifting of the refinement
proof, which, we believe, lowers the proof burden dramatically.
Another difference is that their notion of concurrent certified abstraction
layers is more complex than our models, which are plain LTSs; but their models
can be composed together.


\paragraph{Paxos Verification Efforts}
Paxos and its multiple variants have been considered by many verification efforts using, \eg{}, automated theorem provers and model checkers \cite{DBLP:journals/pacmpl/PadonLSS17, DBLP:conf/cav/MaricSB17, DBLP:conf/fm/ChandLS16, DBLP:journals/afp/JaskelioffM05, kellomaki2004}.
These efforts all consider abstract \emph{models} or specifications in high-level domain-specific languages of Paxos(-like) protocols and not actual implementations in a realistic and expressive programming language.

\citet{DBLP:conf/pldi/KraglEHMQ20} work in the Boogie verifier and programming language where they express their high-level model, their low-level (non-distributed) imperative implementation, and the layers in between, all in Boogie.
By contrast, our effort is foundational (in the technical sense of being formalized in Coq as mentioned above) and the implementation is carried out in the OCaml-like programming language \AnerisLang{} with UDP network primitives.

\citet{DBLP:conf/esop/Garcia-PerezGMS18} devise composable specifications for a pseudo-code implementation of Single-Decree Paxos and semantics-preserving optimizations to the protocol on pen-and-paper but without a formal connection to their implementation in Scala; it would be interesting future work to implement and verify the same optimizations in our setting.

\section{Conclusion and Future Work}
\label{sec:conclusion}

In this paper, we explored how intensional refinement \emph{is} indeed a viable methodology for strengthening higher-order concurrent and distributed separation logics to non-trivial safety and liveness properties using \Trillium{} and its instantiations.
We have developed \Fairis{}, a higher-order concurrent separation logic, and we have shown how the logic gives us a methodology for proving liveness of concurrent programs under fair scheduling assumptions.
Moreover, we instantiated \Trillium{} with a distributed language and obtained an extension of \Aneris{}, a distributed separation logic, that we have used to show refinement relations between distributed systems and their \TLA{} models.

Future work includes extending \Trillium's support for modular reasoning to also allow specifications of library functions to be modular with respect to the model, such that a library function can be specified with respect to one model and client code can be specified with respect to another model in isolation.
Currently, library functions can be reasoned about modularly in \Trillium{} using higher-order specifications.
For example, our Paxos implementation makes a call to the $\paxosSendtoall$ function, whose specification quantifies over arbitrary socket protocols (themselves higher-order predicates), and in turn the $\paxosSendtoall$ function internally makes a call to a function $\paxosSetiter$, whose specification uses impredicative quantification to support arbitrary callback functions.
An important point here, however, is that these library specifications do not interact with the model, \ie{}, execution corresponds to stuttering steps on the model side.
A concrete goal would be to give a modular specification of a fair lock and then verify termination of a client that uses the lock, but only by relying on its specification.

\paragraph{Data availability statement}
The Coq formalization accompanying this work is available on Zenodo
\citep{artifact} and on GitHub at \url{https://github.com/logsem/trillium}.


\begin{acks}                            
  We would like to thank the anonymous reviewers for their valuable remarks and insightful comments which have improved the presentation of this work.

  This work was supported in part by a Villum Investigator grant (no.
  25804), Center for Basic Research in Program Verification (CPV), from the VILLUM Foundation.
  This work has received funding from the European Research Council (ERC) under
  the European Union’s Horizon 2020 research and innovation program (grant
  agreement No. 101003349).
\end{acks}

\bibliography{paper}



\end{document}